\documentclass[sigconf, nonacm]{acmart}
\usepackage[conf={restate,no link to proof}]{proof-at-the-end}
\usepackage{comment}
\usepackage{amsthm}
\usepackage{xcolor}
\usepackage[linesnumbered,ruled,vlined]{algorithm2e}
\usepackage{etoolbox}
\newbool{ARXIV}
\booltrue{ARXIV}
\usepackage{enumitem}

\newtheoremstyle{mystyle}
  {}
  {}
  {\itshape}
  {}
  {\scshape}
  {.}
  { }
  {}
\theoremstyle{mystyle}

\newtheorem{theorem}{Theorem}
\newtheorem{definition}{Definition}
\newtheorem{problem}{Problem}

\DeclareMathOperator{\profile}{prof}
\DeclareMathOperator{\softmax}{sm}

\let\vec\mathbf  
\let\mat\mathbf  

\AtBeginDocument{%
  \providecommand\BibTeX{{%
    \normalfont B\kern-0.5em{\scshape i\kern-0.25em b}\kern-0.8em\TeX}}}

\def\algoname{SM-II\xspace}
\def\gmname{IP\xspace}
\def\wcaname{wCA-I\xspace}
\def\wcainame{wCA-II\xspace}
\newcommand{\addcolor}[1]{\textcolor{black}{#1}}

\begin{document}
\title{Optimizing Fitness-For-Use of Differentially Private Linear Queries}

\newcommand\vldbdoi{10.14778/3467861.3467864}
\newcommand\vldbpages{XXX-XXX}
\newcommand\vldbvolume{14}
\newcommand\vldbissue{10}
\newcommand\vldbyear{2021}
\newcommand\vldbauthors{\authors}
\newcommand\vldbtitle{\shorttitle} 
\newcommand\vldbavailabilityurl{}
\newcommand\vldbpagestyle{empty} 

\author{Yingtai Xiao, Zeyu Ding, Yuxin Wang, Danfeng Zhang, Daniel Kifer}
\affiliation{%
  \institution{Pennsylvania State University}
}
\email{{yxx5224, dxd437, ykw5163, dbz5017, duk17}@psu.edu}

%
%
%
%

\begin{abstract}
    In practice, differentially private data releases are designed to support a variety of applications. A data release is fit for use if it meets target accuracy requirements for each application. 
In this paper, we consider the problem of answering linear queries under differential privacy subject to per-query accuracy constraints. Existing practical frameworks like the matrix mechanism do not provide such fine-grained control (they optimize \emph{total} error, which allows some query answers to be more accurate than necessary, at the expense of other queries that become no longer useful). Thus, we design a fitness-for-use strategy that adds privacy-preserving Gaussian noise to query answers. The covariance structure of the noise is optimized to meet the fine-grained accuracy requirements while minimizing the cost to privacy.

\end{abstract}

\maketitle

\pagestyle{\vldbpagestyle}
\begingroup\small\noindent\raggedright\textbf{PVLDB Reference Format:}\\
\vldbauthors. \vldbtitle. PVLDB, \vldbvolume(\vldbissue): \vldbpages, \vldbyear.\\
\href{https://doi.org/\vldbdoi}{doi:\vldbdoi}
\endgroup
\begingroup
\renewcommand\thefootnote{}\footnote{\noindent
This work is licensed under the Creative Commons BY-NC-ND 4.0 International License. Visit \url{https://creativecommons.org/licenses/by-nc-nd/4.0/} to view a copy of this license. For any use beyond those covered by this license, obtain permission by emailing \href{mailto:info@vldb.org}{info@vldb.org}. Copyright is held by the owner/author(s). Publication rights licensed to the VLDB Endowment. \\
\raggedright Proceedings of the VLDB Endowment, Vol. \vldbvolume, No. \vldbissue\ %
ISSN 2150-8097. \\
\href{https://doi.org/\vldbdoi}{doi:\vldbdoi} \\
}\addtocounter{footnote}{-1}\endgroup

\ifdefempty{\vldbavailabilityurl}{}{
\vspace{.3cm}
\begingroup\small\noindent\raggedright\textbf{PVLDB Artifact Availability:}\\
The source code, data, and/or other artifacts have been made available at \url{\vldbavailabilityurl}.
\endgroup
}



\section{Introduction}\label{sec:intro}
Differential privacy gives data collectors the ability to publish information about sensitive datasets while protecting the confidentiality of the users who supplied the data. Real-world applications include OnTheMap \cite{onthemap,ashwin08:map}, Yahoo Password Frequency Lists \cite{BlockiDB16}, Facebook URLs Data \cite{FBUrl}, and the 2020 Decennial Census of Population and Housing \cite{abowd}.
Lessons learned from these early applications help identify deployment challenges that should serve as guides for future research. One of these challenges is supporting applications that end-users care about \cite{garfinkel18}. It is well-known that no dataset can support arbitrary applications while providing a meaningful degree of privacy \cite{dinur:privacy}. Consequently, as shown in theory and in practice, privacy-preserving data releases must be carefully designed with intended use-cases in mind. Without such use-cases, a so-called ``general-purpose'' data release might not provide sufficient accuracy for any practical purpose.

Ensuring accuracy for pre-specified use-cases has strong precedents and was common practice even before the adoption of differential privacy. For example, in the 2010 Decennial Census, the U.S. Census Bureau released data products such as:
\begin{itemize}[leftmargin=*]
    \item PL 94-171 Summary Files \cite{pl94} designed to support redistricting. The information was limited to the total number of people, various race/ethnicity combinations in each Census block, along with the number of such people  with age 18 or higher.
    \item Advance Group Quarters Summary File \cite{agq} containing the number of people living in group quarters such as correctional institutions, university housing, and military quarters. One of its purposes is to help different states comply with their own redistricting laws regarding prisoners \cite{grovesblog}.
    \item Summary File 1 \cite{census2010} provides a fixed set of tables that are commonly used to allocate federal funds and to support certain types of social science research.
\end{itemize}

Thus we consider a setting where a data publisher must release differentially private query answers to support a given set of $N$ applications. Each application provides measures of accuracy that, if met, make the differentially private query answers fit for use. \addcolor{For example, one of the measures used by the American Community Survey (ACS) is \emph{margin of error} \cite{aschandbook}, an estimated confidence interval that is a function of variance.}

In this paper, we study the case where the workload consists of linear queries $\vec{w}_1,\dots, \vec{w}_N$. Differential privacy does not allow exact query answers to be released, so the data publisher must release noisy query answers instead. We consider the following fitness-for-use criteria: each workload query $\vec{w}_i$ must be answered with expected squared error $\leq c_i$ (where $c_1,\dots, c_N$ are user-specified constants \addcolor{that serve as upper bounds on desired error}). 

Given these fitness-for-use constraints, the data publisher must determine whether they can be met under a given privacy budget and, if so, how to correlate the noise in the query answers in order to meet these constraints.\footnote{If the desired accuracy cannot be met under a given privacy budget, the data collector could reinterpret the constants $c_i$ to represent relative priorities, see Section \ref{sec:problem}.}
We note that earlier applied work on the matrix mechanism \cite{YYZH16,LMHMR15,YZWXYH12} optimized \emph{total} error rather than per-query error, and could not guarantee that each query is fit for use. Theoretical work on the matrix mechanism \cite{edmonds2020power,nikolov2014new} studied expected worst-case, instead of per-query, error and, as we discuss later, it turns out that the same algorithm can optimize them.

We propose a mechanism that adds correlated Gaussian noise to the query answers. The correlation structure is designed to meet accuracy constraints while minimizing the privacy cost, and is obtained by solving an optimization problem. We analyze the solution space and show that although there potentially exist many such correlation structures, there is a unique solution that allows the maximal release of additional  noisy query answers \emph{for free} -- that is, some queries that cannot be derived from the differentially private workload answers can be noisily answered without affecting the privacy parameters \addcolor{(this is related to personalized privacy \cite{personaldp}).}

\noindent Our contributions are:
\begin{itemize}[leftmargin=*]
    \item A novel differentially private mechanism for releasing linear query answers subject to fitness-for-use constraints (to the best of our knowledge, this is the first such mechanism). It uses non-trivial algorithms for optimizing the covariance matrix of the Gaussian noise that is added to the query answers.
    \item We theoretically study the fitness-for-use problem. Although there are potentially infinitely many covariance matrices that can be used to minimize privacy cost while meeting the accuracy constraints, we show that there is a unique solution that allows the data publisher to noisily answer a maximal amount of additional queries at no extra privacy cost.
    \item We design experiments motivated by real-world use cases to show the efficacy of our approach.
\end{itemize}

The outline of this paper is as follows. In Section \ref{sec:background}, we present notation and background material. We formalize the problem in Section \ref{sec:problem}. We discuss related work in Section \ref{sec:related}. We present theoretical results of the solution space in Section \ref{sec:theory}. We present our optimization algorithms for the fitness-for-use problem in Section \ref{sec:optimize}. We present experiments in Section \ref{sec:experiments} and conclusions in Section \ref{sec:conc}. \addcolor{All proofs can be found in \ifbool{ARXIV}{the appendix}{the full version \cite{fullfitness} of this paper}.}



\section{Notation and Background}\label{sec:background}
\begin{table}[t]
\begin{center}
\caption{Table of Notation}\label{tab:notation}
\resizebox{\linewidth}{!}{
\begin{tabular}{|cl|}\hline
$\vec{x}$: & Dataset vector \\
$d$:     & \addcolor{Domain size of the data}\\
$\mat{W}$: & Workload query matrix.\\
$m$: & Number of workload queries.\\
$k$: & Rank of $\mat{W}$.\\
$\vec{y}$: &True answers to workload queries ($\vec{y}=\mat{W}\vec{x}$).\\
$\mat{B}$: & Basis matrix with linearly independent rows.\\
$\vec{b}_i$: & The $i^\text{th}$ column of $\mat{B}$.\\
$\mat{L}$: & Representation matrix. Note, $\mat{L}\mat{B}=\mat{W}$.\\
$\vec{c}$: &Vector of accuracy targets ($c_1,\dots, c_m$).\\
$\preceq_{R}$: & Refined privacy ordering (Definition \ref{def:order}).\\
$\profile(\mat{\Sigma},\mat{B}):$ & Privacy profile (Definition \ref{def:profile}).\\
$\Delta_M$: & Privacy cost (Corollary \ref{corr:query}).\\
\hline
\end{tabular}
}
\end{center}
\end{table}

We denote vectors as bold lower-case letters (e.g., $\vec{x}$), matrices as bold upper-case (e.g., $\mat{W}$), scalars as non-bold lower-case (e.g., $c$).

Following earlier work on differentially private linear queries \cite{YYZH16,LMHMR15,YZWXYH12}, we work with a table whose attributes are categorical (or have been discretized). As in prior work, we represent such a table as a vector $\vec{x}$ of counts. That is, letting $\{t_0, t_1, \dots t_{d-1}\}$ be the set of possible tuples,  $\vec{x}[i]$ is the number of times tuple $t_i$ appears  in the table. For example, consider a table on two attributes, \emph{adult} (yes/no) and \emph{Hispanic} (yes/no). We set $t_0=$``not adult, not Hispanic'', $t_1=$``adult, not Hispanic'', $t_2=$``not adult, Hipanic'', $t_3=$``adult, Hispanic''. Then $\vec{x}[3]$ is the number of Hispanic adults in the dataset. 

We refer to $\vec{x}$ as a \emph{dataset vector} and we say that two dataset vectors $\vec{x}$ and $\vec{x}^\prime$ are \emph{neighboring} (denoted as $\vec{x}\sim\vec{x}^\prime$) if $\vec{x}$ can be obtained from $\vec{x}^\prime$ by adding or subtracting 1 from some component of $\vec{x}^\prime$ (this means $||\vec{x}-\vec{x}^\prime||_1=1$) -- this is the same as adding or removing 1 person from the underlying table.

A single linear query $\vec{w}$ is a vector, whose answer is $\vec{w}\cdot\vec{x}$. A set of $m$ linear queries is represented by an $m\times d$ matrix $\mat{W}$, where each row corresponds to a single linear query. The answers to those queries are obtained by matrix multiplication: $\mat{W}\vec{x}$. For example, for the query matrix $W=\left(\begin{smallmatrix}0 & 1 & 0 & 1\\0 & 0 & 1 & 1\\1 & 1 & 1 & 1\end{smallmatrix}\right)$, the first row is the query for number of adults in the table (since $0 \vec{x}[0] + 1\vec{x}[1] + 0 \vec{x}[2] + 1\vec{x}[3]$ sums up over the tuples corresponding to adults); the second row is the query for number of Hispanic individuals; and the last row is the query for the total number of people. We summarize notation in Table \ref{tab:notation}.

Our privacy mechanisms are compatible with many variations of differential privacy, including concentrated differential privacy \cite{zcdp} and Renyi differential privacy \cite{renyidp}. As these are complex definitions, for simplicity we focus on approximate differential privacy \cite{dwork06Calibrating,dworkKMM06:ourdata}, defined as follows.

\begin{definition}[$(\epsilon,\delta)$-Differential Privacy \cite{dworkKMM06:ourdata}]\label{def:dp} Given privacy parameters $\epsilon > 0$ and $\delta \in (0, 1)$, a randomized algorithm $M$ satisfies $(\epsilon,\delta)$-differential privacy if for all pairs of neighboring dataset vectors $\vec{x}$ and $\vec{x}^\prime$ and all sets $S$, the following equations hold:
\begin{align*}
    P(M(\vec{x})\in S) \leq e^{\epsilon} P(M(\vec{x}^\prime)\in S) + \delta
\end{align*}
\end{definition}

Intuitively, differential privacy guarantees that the output distribution of a randomized algorithm is barely affected by any person's record being included in the data.

In the case of privacy-preserving linear queries, this version of differential privacy is commonly achieved by adding independent Gaussian noise to the query answers $\mat{W}\vec{x}$. The scale of the noise depends on the $L_2$ sensitivity of the queries.
\begin{definition}[$L_2$ sensitivity]\label{def:l2sensitivity}
The $L_2$ sensitivity of a function $f$, denoted by $\Delta_2(f)$ is defined as $\sup_{\vec{x}\sim \vec{x}^\prime}||f(\vec{x})-f(\vec{x}^\prime)||_2$.
\end{definition}
If $f$ computes linear query answers (i.e., $f(\vec{x})=\mat{W}\vec{x}$) then we slightly abuse notation and denote the $L_2$ sensitivity as $\Delta_2(W)$. It follows \cite{YYZH16} that $\Delta_2(W)$ is equal to $\max_i || W[:, i]||_2$, where $W[:, i]$ is the $i^\text{th}$ column of $W$.

The Gaussian Mechanism adds independent noise with variance $\sigma^2$  to $f(\vec{x})$ and releases the resulting noisy query answers. Its differential privacy properties are provided by the following theorem.

\begin{theorem}[Exact Gaussian Mechanism \cite{BW18}]\label{thm:gaussmech}
Let $\Phi$ be the cumulative distribution function of the standard Gaussian distribution. Let $f$ be a (vector-valued) function. Let $M$ be the algorithm that, on input $\vec{x}$, outputs $f(\vec{x})+\vec{z}$, where $\vec{z}$ is an $m$-dimensional vector of independent $N(0, \sigma^2)$ random variables. Then $M$ satisfies $(\epsilon,\delta)$-differential privacy if and only if
\begin{align*}
    \delta &\geq \Phi\left(\frac{\Delta_2(f)}{2\sigma}-\frac{\epsilon\sigma}{\Delta_2(f)}\right) - e^{\epsilon}\Phi\left(-\frac{\Delta_2(f)}{2\sigma}-\frac{\epsilon\sigma}{\Delta_2(f)}\right)
\end{align*}
Furthermore, this is an increasing function of $\Delta_2(f)/\sigma$.
\end{theorem}
The importance of this result is that the privacy properties of the Gaussian mechanism are completely determined by $\Delta_2(f)/\sigma$ (in fact, this is true for $(\epsilon,\delta)$-differential privacy, Renyi differential privacy and concentrated differential privacy). In particular, decreasing $\Delta_2(f)/\sigma$ increases the amount of privacy  \cite{BW18}.

\addcolor{\emph{Personalized differential privacy} \cite{personaldp} refines  differential privacy by letting each domain element have its own privacy parameter. For example, a domain element $r$ has privacy parameters $(\epsilon, \delta)$ if  $P(M(\vec{x})\in S) \leq e^{\epsilon} P(M(\vec{x}^\prime)\in S) + \delta$ for all $S$ and neighbors $\vec{x}$ and $\vec{x}^\prime$ that differ by 1 in the count of record $r$.}

\section{The Fitness-for-use Problem \label{sec:prob}}\label{sec:problem}

In this section, we start with an intuitive problem statement and then mathematically formalize it, while justifying our design choices along the way.
We consider two problems: one that prioritizes accuracy and a second that prioritizes privacy. Later we will show that they are equivalent (the same method yields a solution to both).

When \textbf{prioritizing accuracy}, the workload consists of $m$ linear queries  $\vec{w}_1, \dots, \vec{w}_n$. Given a vector $\vec{c}$ of accuracy targets, we seek to find a mechanism $M$ that produces $(\epsilon,\delta)$-differentially private answers to these queries such that the expected squared error for each query $\vec{w}_i$ is less than or equal to $\vec{c}[i]$. The mechanism should maximize privacy subject to these accuracy constraints. Maximizing privacy  turns out to be a surprisingly subtle and nontrivial concept involving (1) the minimization of the  privacy parameters $\epsilon$ and $\delta$ and (2) enabling the data publisher to noisily answer extra queries at a later point in time at no additional privacy cost; it is discussed in detail in Section \ref{subsec:maxprivacy}.
When \textbf{prioritizing privacy}, again we have a workload of $m$ linear queries  $\vec{w}_1, \dots, \vec{w}_n$ and a vector $\vec{c}$. Given target privacy parameters $\epsilon$ and $\delta$, the goal is to find a mechanism $M$ that (1) satisfies $(\epsilon,\delta)$-differential privacy and (2) finds the smallest number $k > 0$ such that each query $\vec{w}_i$ can be privately answered with accuracy at most $k\vec{c}[i]$. Thus $\vec{c}[i]$ represents the relative importance of query $\vec{w}_i$ and the goal is to obtain the most accurate query answers while respecting the privacy constraints and relative accuracy preferences.

We next discuss additional accuracy-related desiderata (Section \ref{subsec:desiderata}), formalize the maximization of privacy (Section \ref{subsec:maxprivacy}), and then fully formalize the fitness-for-use problem (Section \ref{subsec:problem}).

\subsection{Additional Accuracy Desiderata}\label{subsec:desiderata}

We first require that the differentially private mechanism $M$ produces \emph{unbiased} noisy query answers (i.e., the expected value of the noisy query answers equals the true value). Unbiased query answers allow end-users to compute the expected error of derived queries. This property is best explained with an example. Suppose a mechanism $M$ provides  noisy counts (with independent noise) for (1) $\widetilde{y}_1$: the number of adults with cancer, and (2) $\widetilde{y}_2$: the number of children with cancer. From these two queries, we can derive an estimate for the total number of cancer patients as $\widetilde{y}_1+\widetilde{y}_2$. Suppose $\widetilde{y}_1$ has expected squared error $c_1$ and $\widetilde{y}_2$ has expected squared error $c_2$. What can we say about the expected squared error of $\widetilde{y}_1 + \widetilde{y}_2$? If both $\widetilde{y}_1$ and $\widetilde{y}_2$ are unbiased, then the expected squared error of $\widetilde{y}_1 + \widetilde{y}_2$ equals $c_1+c_2$. However, if  $\widetilde{y}_1$ and $\widetilde{y}_2$ are biased, then the expected squared error of $\widetilde{y}_1 + \widetilde{y}_2$ cannot be accurately determined from the expected errors of $\widetilde{y}_1$ and $\widetilde{y}_2$.

Since statistical end-users need to be able to estimate the errors of quantities they derive from noisy workload query answers, \addcolor{we thus require our mechanism to produce unbiased query answers.} 

We next require that the noise added to query answers should be correlated multivariate Gaussian  noise $N(\vec{0}, \mat{\Sigma})$. Gaussian noise is familiar to statisticians and simplifies their calculations for tasks such as performing hypothesis tests and computing confidence intervals \cite{Wang2019:DPC}. Furthermore, Gaussian noise is often a preferred distribution for various versions of approximate differential privacy \cite{deepdp,renyidp,zcdp,diffpbook}.

For readers familiar with the matrix mechanism \cite{YYZH16,LMHMR15,YZWXYH12} (which optimizes total squared error, not per-query error), it is important to note that we add correlated noise in a different way. The matrix mechanism starts with a workload of linear queries $\mat{W}$ and solves an optimization problem to get a different collection $\mat{S}$ of linear queries, called the \emph{strategy matrix}. Noisy strategy answers $\widetilde{\vec{y}}_s$ are obtained by adding independent noise $\vec{z}$ to the strategy queries ($\widetilde{\vec{y}}_s = \mat{S}\vec{x} + \vec{z}$) and the workload query answers $\widetilde{\vec{y}}_w$ are reconstructed as follows $\widetilde{\vec{y}}_w = \mat{W}\mat{S}^+\widetilde{\vec{y}}_s$ (where $\mat{S}^+$ is the Moore-Penrose pseudo-inverse of $\mat{S}$ \cite{GoluVanl96}).
Instead of optimizing for a matrix $\mat{S}$ and adding independent Gaussian noise, we fix the matrix and optimize the correlation structure of the noise (as in \cite{nikolov2014new}). This turns out to be a more convenient formulation for our problem. Formally,\footnote{In \cite{edmonds2020power}, this is called the factorization mechanism. For their metric, they propose setting $\mat{\Sigma}=\mat{I}$  while optimizing $\mat{B}$ and $\mat{L}$. In our work, it is easier to optimize $\mat{\Sigma}$ instead.}
\begin{definition}[Linear Query Mechanism \cite{edmonds2020power}]\label{def:basic}
Let $\mat{W}$ be a linear query workload matrix. Let $\mat{B}$ and $\mat{L}$ be matrices such that $\mat{W}=\mat{L}\mat{B}$ and $\mat{B}$ has linearly independent rows with the same rank as $\mat{W}$ (we call $\mat{B}$ the \emph{basis} matrix and $\mat{L}$ the \emph{representation} matrix). Let $N(\vec{0}, \mat{\Sigma})$ be the multivariate Gaussian distribution with covariance matrix $\Sigma$. Then the mechanism $M$, on input $\vec{x}$, outputs $\mat{L}(\mat{B}\vec{x} + N(\vec{0}, \mat{\Sigma}))$.
\end{definition}

\addcolor{Note that $E[\mat{L}(\mat{B}\vec{x} + N(\vec{0}, \mat{\Sigma}))]=\mat{L}\mat{B}\vec{x}=\mat{W}\vec{x}$ so the mechanism is indeed unbiased.} 
The basis matrix $\mat{B}$ is chosen to be any convenient matrix after the workload is known (but before the dataset is seen). It is there for computational convenience -- a linear query mechanism can be represented using any basis matrix we want, as shown in Theorem \ref{thm:basis}. Hence we can choose the basis  $\mat{B}$ (and corresponding $\mat{L}$) for which we can speed up matrix operations in our optimization algorithms. In our work, we typically set $\mat{B}$ to be the identity matrix or a linearly independent subset of the workload matrix. 

%
%
\begin{theoremEnd}[category=section3]{theorem}\label{thm:basis}
Given a workload $\mat{W}$ and two possible decompositions: $\mat{W}=\mat{L}_1\mat{B}_1$ and $\mat{W}=\mat{L}_2\mat{B}_2$, where the rows of $\mat{B}_1$ are linearly independent with the same rank as $\mat{W}$ (and same for $\mat{B}_2$). 
Let $M_1(\vec{x})=\mat{L}_1(\mat{B}_1\vec{x}+N(\vec{0},\mat{\Sigma}_1))$. There exists a $\mat{\Sigma}_2$ such that the mechanism $M_2(\vec{x})=\mat{L}_2(\mat{B}_2\vec{x}+N(\vec{0},\mat{\Sigma}_2))$ has the same output distribution as $M_1$ for all $\vec{x}$.
\end{theoremEnd}
\begin{proofEnd}
Since $\mat{B}_1$ has linearly independent rows, it has a right inverse, which we denote as $\mat{B}_1^{-R}$. Since $\mat{W}=\mat{L}_1\mat{B}_1$, this means $\mat{L}_1=\mat{W}\mat{B}_1^{-R}$. Similarly, let $\mat{B}_2^{-R}$ be the right inverse of $\mat{B}_2$, so  $\mat{L}_2=\mat{W}\mat{B}_2^{-R}$.

Define $\mat{\Sigma}_2=\mat{B}_2\mat{B}_1^{-R}\mat{\Sigma}_1 (\mat{B}_1^{-R})^T \mat{B}_2^T$. We will use the following facts:
\begin{description}
    \item[Fact 1:] $\mat{W}\mat{B}_2^{-R}\mat{B}_2=\mat{W}$, this follows from $\mat{W}=\mat{L}_2\mat{B}_2$ and $\mat{L}_2=\mat{W}\mat{B}_2^{-R}$ (i.e., in a limited sense, $\mat{B}_2^{-R}$ can sometimes act like a left inverse for $\mat{B}$, since $\mat{W}$ is in the row space of $\mat{B}_2$).
    \item[Fact 2:] If $\vec{z}$ is a $N(\vec{0}, \mat{\Sigma}_2)$ random variable then $\mat{L}_2\vec{z}$ is a $N(\vec{0}, \mat{L}_2\mat{\Sigma}_2\mat{L}_2^T)$ random variable.
\end{description}
Then it follows that:
\begin{align*}
    M_2(\vec{x})&= \mat{L}_2(\mat{B}_2\vec{x}+N(\vec{0},\mat{\Sigma}_2))\\
                &= \mat{W}\vec{x} +  \mat{L}_2N(\vec{0},\mat{\Sigma}_2))\\
                &= \mat{W}\vec{x} + N(\vec{0},\mat{L}_2\mat{\Sigma}_2\mat{L}_2^T)\\
                &= \mat{W}\vec{x} + N(\vec{0},\mat{W}\mat{B}_2^{-R}\mat{\Sigma}_2(\mat{B}_2^{-R})^T\mat{W}^T)\\
                &= \mat{W}\vec{x} + N(\vec{0},\mat{W}\mat{B}_2^{-R}\mat{B}_2\mat{B}_1^{-R}\Sigma (\mat{B}_1^{-R})^T \mat{B}_2^T(\mat{B}_2^{-R})^T\mat{W}^T)\\
                &= \mat{W}\vec{x} + N(\vec{0}, \mat{W}\mat{B}_1^{-R}\Sigma (\mat{B}_1^{-R})^T \mat{W}^T)\\
                &= \mat{W}\vec{x} + N(\vec{0}, \mat{L}_1\Sigma \mat{L}_1^T)\\
                &= \mat{W}\vec{x} + \mat{L}_1 N(\vec{0}, \Sigma )\\
                &= \mat{L}_1(\mat{B}_1\vec{x}+N(\vec{0},\mat{\Sigma}_1))\\
                &= M_1(\vec{x})
\end{align*}
\end{proofEnd}

\subsection{Maximizing Privacy}\label{subsec:maxprivacy}
In this section we explain how to compare the privacy properties of the basic mechanisms from Definition \ref{def:basic}, based on their privacy parameters and additional noisy query answers they can release for free. We first generalize Theorem \ref{thm:gaussmech} to allow correlated noise. 
\begin{theoremEnd}[category=section3]{theorem}[Exact Correlated Gaussian Mechanism]\label{thm:corrgaussian}
Let $f$ be a function that returns a vector of query answers. Let $M$ be the mechanism that, on input $\vec{x}$, outputs $f(\vec{x}) + N(\vec{0}, \mat{\Sigma})$. Define the quantity
\begin{align}
\Delta_{\Sigma}(f)=\max_{\vec{x}_1\sim \vec{x}_2} \sqrt{\left|(f(\vec{x}_1)-f(\vec{x}_2))^T \mat{\Sigma}^{-1} (f(\vec{x}_1)-f(\vec{x}_2))\right|}\label{eqn:correlated_sensitivity}
\end{align}
where the max is over all pairs of neighboring datasets $\vec{x}_1$ and $\vec{x}_2$. Then $M$ satisfies $(\epsilon,\delta)$-differential privacy if and only if:
\begin{align}
    \delta &\geq \Phi\left(\frac{\Delta_{\mat{\Sigma}}(f)}{2}-\frac{\epsilon}{\Delta_{\mat{\Sigma}}(f)}\right) - e^{\epsilon}\Phi\left(-\frac{\Delta_{\mat{\Sigma}}(f)}{2}-\frac{\epsilon}{\Delta_{\mat{\Sigma}}(f)}\right)\label{eqn:correlated_delta}
\end{align}
where $\Phi$ is the CDF of the standard Gaussian $N(0,1)$. Furthermore, this quantity is an increasing function of ${\Delta_\Sigma(f)}$.
\end{theoremEnd}
\begin{proofEnd}
Let $\mat{\Sigma}^{1/2}$ denote the matrix square root of $\mat{\Sigma}$ (i.e., $\mat{\Sigma}^{1/2}\mat{\Sigma}^{1/2}=\mat{\Sigma}$), which exists because $\mat{\Sigma}$ is a covariance matrix and hence symmetric positive definite. Define the function $g(\vec{x}) = \mat{\Sigma}^{-1/2}f(\vec{x})$ and  $M^\prime(\vec{x})=g(\vec{x}) + N(\vec{0}, \mat{I})$. Note that $M(\vec{x})=\mat{\Sigma}^{1/2}M^\prime(\vec{x})$  (that is, $M$ is equivalent to first running the mechanism $M^\prime$ on $\vec{x}$ and then multiplying the result by $\mat{\Sigma}^{1/2}$). Since $\mat{\Sigma}$ is invertible, then by the postprocessing theorem of differential privacy \cite{diffpbook}, $M$ and $M^\prime$ have the exact same privacy parameters. $M^\prime$ is the Gaussian mechanism applied to the function $g$, whose $L_2$ sensitivity  is
\begin{align*}
    \Delta_2(g) &=\addcolor{\max_{\vec{x}_1\sim \vec{x}_2}||g(\vec{x}_1) - g(\vec{x}_2)||_2}  \\
    &= \max_{\vec{x}_1\sim\vec{x}_2}||\mat{\Sigma}^{-1/2}f(\vec{x}_1) - \mat{\Sigma}^{-1/2}f(\vec{x}_2) ||_2 \\  &= \sqrt{(f(\vec{x}_1) - f(\vec{x}_2)^T\mat{\Sigma}^{-1}(f(\vec{x}_1) - f(\vec{x}_2))} \\ &=\Delta_{\mat{\Sigma}}(f)
\end{align*}
 Applying Theorem \ref{thm:gaussmech} to $M^\prime$ and substituting $\Delta_{\mat{\Sigma}}(f)$ in place of $\Delta_2(g)$, we see that $M^\prime$ (and hence $M$) satisfies $(\epsilon,\delta)$-differential privacy if and only Equation \ref{eqn:correlated_delta} is true.
\end{proofEnd}

Applying Theorem \ref{thm:corrgaussian} to the linear query mechanisms, we get:
\begin{theoremEnd}[category=section3]{corollary}\label{corr:query}
Let $M$ be a linear query mechanism (Definition \ref{def:basic}) with basis matrix $\mat{B}$ and Gaussian covariance $\mat{\Sigma}$. Let $\vec{b}_1,\dots, \vec{b}_d$ be the columns of $\mat{B}$ and denote the privacy cost $\Delta_M=\max_{i=1,\dots, d} $  $ \sqrt{\vec{b}_i^T\mat{\Sigma}^{-1}\vec{b}_i}$. Then $M$ satisfies $(\epsilon,\delta)$-differential privacy if and only if $\delta\geq \Phi\left(\frac{\Delta_M}{2}-\frac{\epsilon}{\Delta_M}\right) - e^{\epsilon}\Phi\left(-\frac{\Delta_M}{2}-\frac{\epsilon}{\Delta_M}\right)$. Furthermore, this quantity is increasing in $\Delta_M$.
\end{theoremEnd}
\begin{proofEnd}
From Definition \ref{def:basic}, recall that a linear query mechanism $M$ uses 4 matrices: $\mat{W}, \mat{B}, \mat{L}, \mat{\Sigma}$. $\mat{W}$ is the $m\times d$ workload matrix. Let $k$ be the rank of $\mat{W}$. Furthermore, recall $\mat{W}=\mat{L}\mat{B}$, and that $\mat{B}$ has linearly independent rows and the same rank as $\mat{W}$ so that $\mat{B}$ is a $k\times d$ matrix. This means that the $m\times k$ matrix $\mat{L}$ also has the same rank as $\mat{W}$ and hence has rank $k$ (same as the number of columns of $\mat{L}$). Therefore $\mat{L}$ has a left inverse $\mat{L}^{-\text{left}}$.

Consider the mechanism $M^\prime$ defined as $M^\prime(\vec{x})=L^{-\text{left}}M(\vec{x}) = \mat{B}\vec{x} + N(\vec{0}, \mat{\Sigma})$. By the postprocessing theorem of differential privacy \cite{diffpbook}, $M^\prime$ and $M$ satisfy $(\epsilon,\delta)$-differential privacy for the exact same value of $\epsilon$ and $\delta$. Clearly, $M^\prime$ is the correlated Gaussian Mechanism applied to the function $f(\vec{x})=\mat{B}(\vec{x})$. Using the notation that $\vec{e}_i$ is the vector that has a 1 in position $i$ and 0 everywhere else, we compute $\Delta_{\mat{\Sigma}}(f)$ as follows:
\begin{align*}
    \Delta_{\mat{\Sigma}}(f) &= \max_{\vec{x}_1\sim\vec{x}_2} \sqrt{(f(\vec{x}_1)-f(\vec{x}_2)^T\mat{\Sigma}^{-1}(f(\vec{x}_1)-f(\vec{x}_2))}\\
                    &=  \max_{\vec{x}_1\sim\vec{x}_2} \sqrt{(B\vec{x}_1- B\vec{x}_2)^T\mat{\Sigma}^{-1}(B\vec{x}_1-B\vec{x}_2)}\\
                    &= \max_{i=1,dots, d} \sqrt{(B\vec{e}_i)^T\mat{\Sigma}^{-1}(B\vec{e}_i)}\\
                    &= \max_{i=1,\dots, d} \sqrt{\vec{b}_i^T\mat{\Sigma}^{-1}\vec{b}_i}
\end{align*}
Theorem \ref{thm:corrgaussian} to $M^\prime$ now yields the result.
\end{proofEnd}

\begin{figure}
    \centering
    \includegraphics[width=0.40\textwidth]{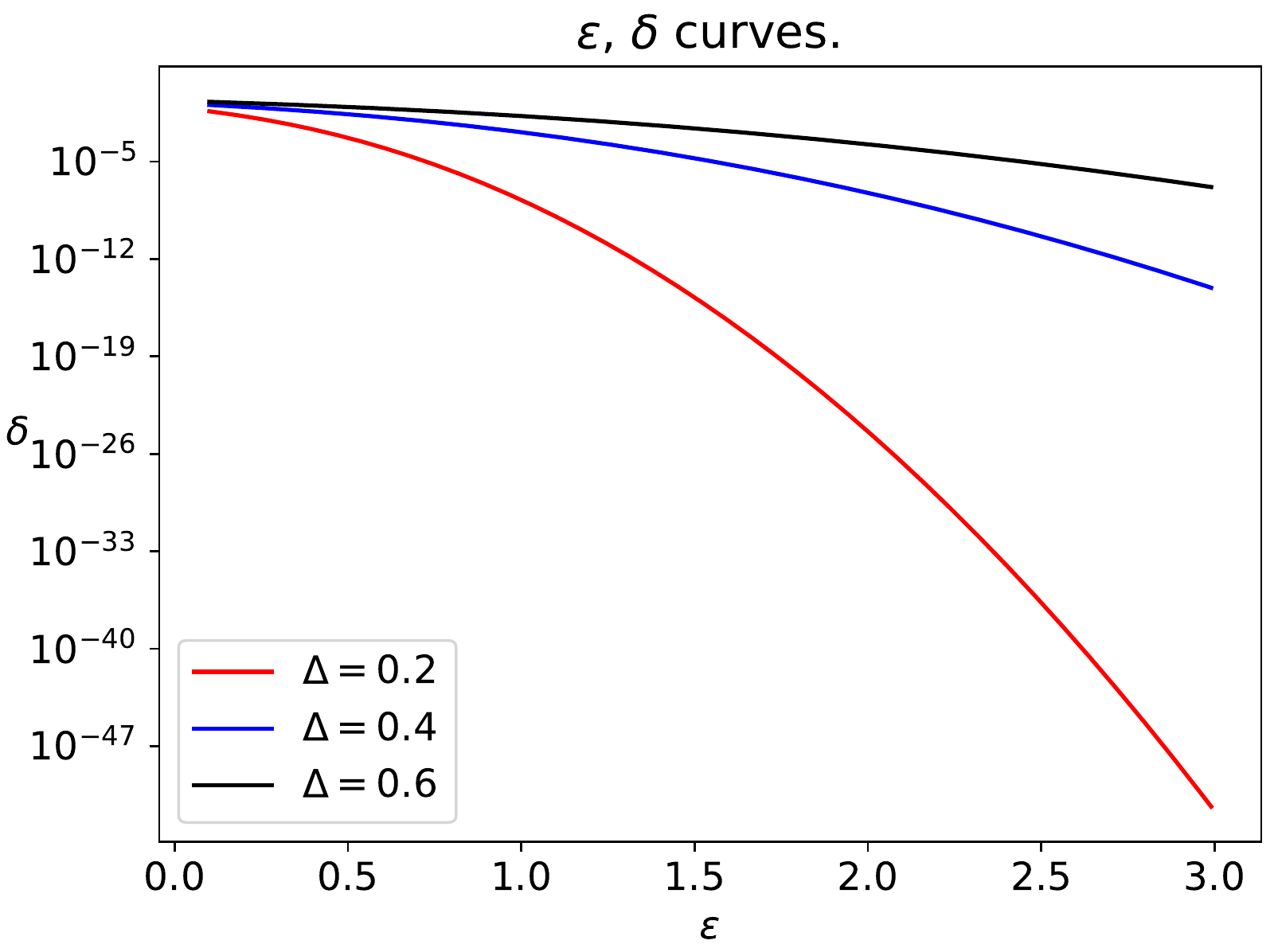}
    \caption{$\epsilon$, $\delta$ curves for different values of $\Delta$. A Gaussian Mechanism with $\Delta=0.2$ satisfies $(\epsilon,\delta)$-differential privacy for all pairs $(\epsilon,\delta)$ that lie on or above the red curve.}
    \label{fig:curve}
\end{figure}

Any given mechanism $M$ typically satisfies $(\epsilon,\delta)$-differential privacy for infinitely many $\epsilon, \delta$ pairs, which defines a curve in space  \cite{renyidp} (see Figure \ref{fig:curve}).
The importance of Corollary $\ref{corr:query}$ is that this curve for a linear query mechanism $M$ is completely determined by the single number $\Delta_M$ (as defined in Corollary \ref{corr:query}). Furthermore, for any two linear query mechanism $M_1(\vec{x}) = \mat{L}_1(\mat{B}_1\vec{x} + N(\vec{0},\mat{\Sigma}_1))$ and $M_2(\vec{x}) = \mat{L}_2(\mat{B}_2\vec{x} + N(\vec{0},\mat{\Sigma}_2))$, if $\Delta_{M_1} < \Delta_{M_2}$ then the $(\epsilon,\delta)$ curve for $M_1$ is strictly below the curve for $M_2$. This means that the set of pairs $(\epsilon, \delta)$ for which $M_1$ satisfies differential privacy is a strict superset of the pairs for which $M_2$ satisfies differential privacy (while if $\Delta_{M_1}=\Delta_{M_2}$ then the $(\epsilon, \delta)$ curves are exactly the same). For this reason, we call $\Delta_M$ the \textbf{privacy cost}.\footnote{\addcolor{Readers familiar with $\rho$-zCDP \cite{zcdp} should note that privacy cost is the same as $\sqrt{2\rho}$.}}

Thus one goal for maximizing privacy is to choose mechanisms $M$ with as small $\Delta_M$ as possible as this will result in the mechanism satisfying differential privacy with the smallest choices $\epsilon$ and $\delta$ parameters.

\subsubsection{Query answers for Free}
\addcolor{Even for two mechanisms that satisfy differential privacy with exactly the same $\epsilon,\delta$ parameters, it is still possible to say that one provides more privacy than the other by comparing them in terms of personalized differential privacy \cite{personaldp}.} To see why, consider the matrices $\mat{B}_1 = \left(\begin{smallmatrix}1 & 1 & 0\\ 0 & 1 & 1\end{smallmatrix}\right)$ and $\mat{B}_2=\left(\begin{smallmatrix}1 & 1 & 0\\ 0 & 1 & 1\\1 & 0 & 1\end{smallmatrix}\right)$,  and the two mechanisms $M_1(\vec{x}) = \mat{B}_1\vec{x} + N(\vec{0}, \mat{\Sigma}_1)$ and \addcolor{$M_2=\mat{B}_{2}\vec{x} + N(\vec{0}, \mat{\Sigma}_2)$}, where $\mat{\Sigma}_1$ and $\mat{\Sigma}_2$ are identity matrices. Note that the difference between $M_1$ and  $M_2$ is that $M_2$ answers the same queries as $M_1$ plus an additional query (corresponding to the third row of $\mat{B}_2$). However, from Corollary \ref{corr:query}, $\Delta_{M_1}=\Delta_{M_2}$ which means that they satisfy differential privacy for exactly the same privacy parameters.
\addcolor{But, in terms of personalized differential privacy, the personalized privacy cost of domain element $x[i]$ under mechanism $M_1$ is the square root of the $i^\text{th}$ diagonal element of $\mat{B}_1^T\mat{\Sigma}_1^{-1}\mat{B}_1$ (and similarly for $M_2$), while the overall non-personalized privacy cost is the max of these (as in Corollary \ref{corr:query}). The personalized privacy costs of $x[0]$ and $x[2]$ under $M_1$ are smaller than under $M_2$, while the privacy cost of $x[1]$ is the same under $M_1$ and $M_2$. Thus it makes sense to say that $M_2$ is also less private.}\footnote{\addcolor{Note that with uncorrelated noise, the personalized cost of domain element $x[i]$ degenerates to the $L_2$ norm of the $i^\text{th}$ column of the basis matrix. If different columns had different norms, prior work on the matrix mechanism added additional queries until the norms were the same (e.g., using $M_2$ instead of $M_1$ in our example above). We believe this practice should be re-examined -- if $M_1$ satisfies the utility goals, why force the analyst to commit to an extra query and limiting their future options?}}
At the same time, we can say that $M_1$ provides more flexibility to the data publisher than $M_2$. There are many possible choices of additional rows (queries) to add to $\mat{B}_1$ without affecting $\Delta_{M_1}$. \addcolor{In fact, these queries can be determined (and noisily answered) at a later date after noisy answers to the first two queries in $\mat{B}_1$ are released.}

This discussion leads to the concept of a privacy profile \addcolor{(which captures the personalized privacy costs of the domain elements)} and refined privacy ordering \addcolor{for linear query mechanisms.}

\begin{definition}[Privacy Profile]
\label{def:profile}
Given a Linear Query Mechanism $M(\vec{x}) = \mat{L}(\mat{B}\vec{x} + N(\vec{0}, \mat{\Sigma}))$, the privacy profile of $M$ is the $d$-dimensional vector $[\vec{b}_1^T\mat{\Sigma}^{-1}\vec{b}_1, \dots, \vec{b}_d^T\mat{\Sigma}^{-1}\vec{b}_d]$ (where $\vec{b}_i$ is the $i^\text{th}$ column of $\mat{B}$) and is denoted by $\profile(\mat{\Sigma}, \mat{B})$ or by $\profile(M)$ (a slight abuse of notation).
\end{definition}
For example, the privacy profile of $M_1$ above is $[1, 2, 1]$ while the profile for $M_2$ is $[2, 2, 2]$. Note that for any linear query mechanism $M$, $\Delta_M^2$ is equal to the largest entry in the privacy profile of $M$.

The privacy profile is invariant to the choice of basis matrix $\mat{B}$ as shown in Theorem \ref{thm:invariant}. Thus the privacy profile is an intrinsic privacy property of a linear query mechanism rather than a specific parameterization.
\begin{theoremEnd}[category=section3]{theorem}\label{thm:invariant}
Let $M_1(\vec{x})=\mat{L}_1(\mat{B}_1\vec{x} + N(\vec{0}, \mat{\Sigma}_1)) $ and $M_2(\vec{x})=\mat{L}_2(\mat{B}_2\vec{x} + N(\vec{0}, \mat{\Sigma}_2))$ be two mechanisms that have the same output distribution for each $\vec{x}$ (i.e., they add noise to different basis matrices but achieve the same results). Then $\profile(M_1)=\profile(M_2)$.
\end{theoremEnd}
\begin{proofEnd}
Note that $\mat{L}_1\mat{B}_1=\mat{W}=\mat{L}_2\mat{B}_2$.
Also, $\mat{W}$ is an $m\times d$ matrix with rank $k$, and  $\mat{L}_1$ (resp., $\mat{L}_2$) is therefore a full-column rank $m\times k$ matrix and $\mat{B}_1$ (resp., $\mat{B}_2$) is a full-row rank $k\times d$ matrix. This means that $\mat{L}_1$ and $\mat{L}_2$ have left inverses, which we denote as the $k\times m$ matrices $\mat{L}_1^{-L}$ and $\mat{L}_2^{-L}$.

Furthermore, since $M_1(\vec{x})=\mat{L}_1(\mat{B}_1\vec{x} + N(\vec{0}, \mat{\Sigma}_1)) $ and $M_2(\vec{x})=\mat{L}_2(\mat{B}_\vec{x} + N(\vec{0}, \mat{\Sigma}_2))$ have the same distribution, the same holds if we premultiply by $\mat{L}_1^{-L}$ and so $\mat{B}_1\vec{x} + N(\vec{0}, \mat{\Sigma}_1)$ has the same distribution as 
\begin{align*}
    &\mat{L}_1^{-L}\mat{L}_2(\mat{B}_\vec{x} + N(\vec{0}, \mat{\Sigma}_2)) 
    \\
    =& \mat{L}_1^{-L}\mat{L}_2\mat{B}_\vec{x} + N(\vec{0}, ( \mat{L}_1^{-L}\mat{L}_2)\Sigma_2 ( \mat{L}_1^{-L}\mat{L}_2)^T)
\end{align*}
which means that $\mat{\Sigma}_1=( \mat{L}_1^{-L}\mat{L}_2)\Sigma_2 ( \mat{L}_1^{-L}\mat{L}_2)^T$. Furthermore, since $\Sigma_1$ and $\Sigma_2$ are non-singular $k\times k$ matrices, then  the $k\times k$ matrix $\mat{L}_1^{-L}\mat{L}_2$ must be non-singular and hence invertible. Hence $\mat{\Sigma}_1^{-1} = (( \mat{L}_1^{-L}\mat{L}_2)^T)^{-1}\Sigma_2^{-1} ( \mat{L}_1^{-L}\mat{L}_2)^{-1}$

Now, since $\mat{L}_1\mat{B}_1=\mat{L}_2\mat{B}_2$ (i.e. the means of the mechanisms are the same), we have
 $\mat{B}_1=\mat{L}_1^{-1}\mat{L}_2\mat{B}_2$ and so
 \begin{align*}
 \mat{B}_1^T\mat{\Sigma}_1^{-1}\mat{B}_1&=(\mat{L}_1^{-1}\mat{L}_2\mat{B}_2)^T\mat{\Sigma}_1^{-1}(\mat{L}_1^{-1}\mat{L}_2\mat{B}_2)\\
 &=\mat{B}_2^T(\mat{L}_1^{-1}\mat{L}_2)^T\mat{\Sigma}_1^{-1}(\mat{L}_1^{-1}\mat{L}_2)\mat{B}_2\\
 &= \mat{B}^T\mat{\Sigma}_2^{-1}\mat{B}
 \end{align*}
 where the last equality follows from plugging in the relation between $\Sigma_1^{-1}$ and $\Sigma_2^{-1}$. This equality also shows that the privacy profiles are the same.
\end{proofEnd}

Given this invariance, we can use the privacy profile to define a more refined ordering that compares the privacy properties of linear query mechanisms.

\begin{definition}[Refined Privacy Ordering]\label{def:order} Given two Linear Query Mechanisms $M_1$ (with basis $\mat{B}_1$ and covariance matrix $\mat{\Sigma}_1$) and $M_2$ (with basis $\mat{B}_2$ and covariance matrix $\mat{\Sigma}_2$), let $p_1$ be the privacy profile of $M_1$ when sorted in \emph{decreasing} order and let $p_2$ be the sorted privacy profile of $M_2$. We say that $M_1$ is at least as private as $M_2$, denoted by $M_1 \preceq_R M_2$ if  $p_1$ is less than or equal to $p_2$ according to the dictionary (lexicographic) order.  We also denote this as $(\mat{\Sigma}_1, \mat{B}_1) \preceq_R (\mat{\Sigma}_2, \mat{B}_2)$. If the dictionary ordering inequality is strict, we use the notation $\prec_R$.
\end{definition}
Note that if $M$ has sorted privacy profile $p$, then the first element of $p$ is equal to $\Delta_{M_1}^2$. Therefore if $\Delta_{M_1} < \Delta_{M_2}$ then also $M_1 \prec_R M_2$, but if $\Delta_{M_1}=\Delta_{M_2}$, it is still possible that $M_1 \prec_R M$ (hence $\prec_R$ refines our privacy comparisons of mechanisms).
Thus, our goal is to satisfy fitness-for-use constraints while finding the $M$ that is minimal according to $\preceq_R$. This is equivalent to finding an $M$ that minimizes $\Delta_{M}$ (a continuous optimization) and then, among all the mechanisms that fit this criteria, we want to select a mechanism that is minimal according to $\preceq_R$ (a combinatorial optimization because of the sorting performed in the privacy profile). 

We note that $\preceq$ is a weak ordering -- that is, we can have two distinct mechanisms $M_1$ and $M_2$ such that $M_1\preceq_R M_2$ and $M_2\preceq_R M_1$ (i.e., they are two different mechanisms with the same privacy properties). Nevertheless, we will show in Section \ref{sec:theory} that in fact, there exists a unique solution to the fitness-for-use problem.

\subsection{The Formal Fitness-for-use Problem}\label{subsec:problem}
Given a workload $\mat{W}=\mat{L}\mat{B}$ and 
a linear query mechanism $M(\vec{x})=\mat{L}(\mat{B}\vec{x}+N(\vec{0},\mat{\Sigma}))=\mat{W}\vec{x} + \mat{L}N(\vec{0},\mat{\Sigma})$, the variance of the noisy answer to the $i^\text{th}$ query of the workload is the $i^\text{th}$ diagonal element of $\mat{L} \mat{\Sigma} \mat{L}^T$. Thus, we can formalize the ``prioritizing accuracy'' and ``prioritizing privacy'' problems as follows.

\begin{problem}[Prioritizing Accuracy]\label{problem1}
Let $\mat{W}$ be an $m\times d$ workload matrix representing $m$ linear queries. Let $\vec{c}=[c_1, \dots, c_m]$ be accuracy constraints (such that the noisy answer to the $i^\text{th}$ query is required to be unbiased with variance at most $c_i$). Let $\mat{B}$ and $\mat{L}$ be matrices such that $\mat{W}=\mat{L}\mat{B}$ and $\mat{B}$ has linearly independent rows. Select the covariance $\mat{\Sigma}$ of the Gaussian noise by solving the following constrained optimization problem:
\begin{align}
    \alpha,\mat{\Sigma} & \leftarrow \arg\min_{\mat{\alpha,\mat{\Sigma}}} \alpha \label{eq:problem1}\\
     & \phantom{\leftarrow}\text{s.t. } \vec{b}_i^T\mat{\Sigma}^{-1}\vec{b}_i \leq \alpha \text{ for all $i=1,\dots, d$}\nonumber\\
     & \phantom{\leftarrow}\text{and } (diag(\mat{L}\mat{\Sigma} \mat{L}^T))[j] \leq c_j \text{ for $j=1,\dots, m$}\nonumber\\
     & \phantom{\leftarrow}\text{and $\mat{\Sigma}$ is symmetric positive definite.}\nonumber
\end{align}
In case of multiple solutions to Equation \ref{eq:problem1},  choose a $\mat{\Sigma}$ that is minimal under the privacy  ordering $\preceq_R$. Then release the output of the mechanism $M(\vec{x})=\mat{L}(\mat{B}\vec{x} + N(\vec{0}, \mat{\Sigma}))$.
\end{problem}
Note that in Problem \ref{problem1}, $\alpha$ is going to equal the maximum of $\max_{i=1,\dots, d} \vec{b}_i^T\mat{\Sigma}^{-1}\vec{b}_i$ and hence will equal the squared privacy cost, $\Delta^2_M$, for the resulting mechanism (and hence minimizing it will minimize the $\epsilon,\delta$ privacy parameters).

We can also prioritize privacy, following the intuition at the beginning of Section \ref{sec:problem}, as finding the smallest $k$ such that the variance of query $i$ is at most $kc_i$ given a target privacy level.
\begin{problem}[Prioritizing Privacy]\label{problem2} Under the same setting as Problem \ref{problem1}, given a target value $\alpha^*$ for $\Delta_M^2$ (which uniquely defines the $\epsilon,\delta$ curve), solve the following optimization for $\mat{\Sigma}$:
\begin{align}
    k,\mat{\Sigma} & \leftarrow \arg\min_{\mat{k,\mat{\Sigma}}} k \label{eq:problem2}\\
     & \phantom{\leftarrow}\text{s.t. } \vec{b}_i^T\mat{\Sigma}^{-1}\vec{b}_i \leq \alpha^* \text{ for all $i=1,\dots, d$}\nonumber\\
     & \phantom{\leftarrow}\text{and } (diag(\mat{L}\mat{\Sigma} \mat{L}^T))[j] \leq kc_j \text{ for $j=1,\dots, m$}\nonumber\\
     & \phantom{\leftarrow}\text{and $\mat{\Sigma}$ is symmetric positive definite.}\nonumber
\end{align}
In case of multiple solutions to Equation \ref{eq:problem2},  choose a $\mat{\Sigma}$ that is minimal under the refined privacy ordering $\preceq_R$.
\end{problem}

It is interesting to note that we arrived at Problem \ref{problem2} by trying to find a mechanism that minimizes per-query error, which mathematically is the same as minimizing $||\;E[(\mat{W}\vec{x}-M(\vec{x}))^2 ./ \vec{c}] \;||_\infty$ (where $./$ is pointwise division of vectors). Edmonds et al. \cite{edmonds2020power} and Nikolov \cite{nikolov2014new} studied mechanisms that minimize outlier error (also known as joint error): $E[\;||\mat{W}\vec{x}-M(\vec{x}))^2 ./ \vec{c} ||_\infty\;]$. Their nearly-optimal algorithm for this metric can also be obtained by solving Equation \ref{eq:problem2} (which computes an ellipsoid infinity norm \cite{nikolov2014new}) but they did not study the tie-breaking condition in Problem \ref{problem2}.

The following result shows that a solution to Problem \ref{problem1} can be converted into a solution for Problem \ref{problem2}, so for the rest of this paper, we focus on solving the optimization defined in Problem \ref{problem1}.

\begin{theoremEnd}[category=section3]{theorem}\label{thm:equiv}
Let $\mat{\Sigma}$ be a solution to Problem \ref{problem1}. Define the quantities $\alpha=\max_{i=1,\dots,d} \vec{b}_i^T\mat{\Sigma}^{-1}\vec{b}_i$ and   $\mat{\Sigma}_2=\frac{\alpha}{\alpha^*}\mat{\Sigma}$. Then $\mat{\Sigma}_2$ is a solution to Problem \ref{problem2}.
\end{theoremEnd}
\begin{proofEnd}
First note that $\mat{\Sigma}_2$ satisfies the privacy requirements of Problem \ref{problem2} since 
$$\max_{i=1,\dots,d} \vec{b}_i^T\mat{\Sigma}_2^{-1}\vec{b}_i= \max_{i=1,\dots,d} \frac{\alpha^*}{\alpha}\vec{b}_i^T\mat{\Sigma}^{-1}\vec{b}_i=\alpha^*$$
By the optimality of $\mat{\Sigma}$ for Problem \ref{problem1}, we also have for all $i$, $diag(\mat{L}^T\mat{\Sigma}_2 \mat{L})[i] \leq \frac{\alpha}{\alpha^*} c_i$.

Now, by way of contradiction, suppose $\mat{\Sigma}_2$ was not optimal for Problem \ref{problem2}. Then there exists a $\mat{\Sigma}_2^\prime$ such that  $\max_{i=1,\dots,d} \vec{b}_i^T(\mat{\Sigma}_2^\prime)^{-1}\vec{b}_i\leq \alpha^*$ and for all $i$, $diag(\mat{L}^T\mat{\Sigma}_2^\prime \mat{L})[i] \leq k c_i$ for some $k$ and
\begin{description}
    \item[Option 1:] Either $k <  \frac{\alpha}{\alpha^*}$ or
    \item[Option 2:] $k=\frac{\alpha}{\alpha^*}$ but $\profile(\mat{\Sigma}_2^\prime, \mat{B})\preceq_R \profile(\mat{\Sigma}_2, \mat{B})$
\end{description}

If we define $\mat{\Sigma}_1^\prime=\mat{\Sigma}_2^\prime/k$ then this would mean:
\begin{enumerate}
    \item for all $i$, $diag(\mat{L}^T\mat{\Sigma}_1^\prime \mat{L})[i] \leq  c_i$, so $\mat{\Sigma}_1^\prime$ satisfies the accuracy requirements of Problem \ref{problem1}.
    \item $\max_{i=1,\dots,d} \vec{b}_i^T(\mat{\Sigma}_1^\prime)^{-1}\vec{b}_i=\max_{i=1,\dots,d} \vec{b}_i^T(\mat{\Sigma}_2^\prime/k)^{-1}\vec{b}_i\leq k\alpha^*$ 
\end{enumerate}
Now, in the case of Option 1, $k\alpha^* < \alpha$, which means $\mat{\Sigma}_1^\prime$ is a better solution for Problem \ref{problem1} than $\mat{\Sigma}$ (a contradiction).
In the case of Option 2, $k=\frac{\alpha}{\alpha^*}$ and   $\profile(\mat{\Sigma}_1^\prime, \mat{B}) = \frac{1}{k}\profile(\mat{\Sigma}_2^\prime, \mat{B})\prec_R \frac{1}{k}\profile(\mat{\Sigma}_2, \mat{B})=\profile(\mat{\Sigma}, \mat{B})$ which again  means $\mat{\Sigma}_1^\prime$ is a better solution for Problem \ref{problem1} than $\mat{\Sigma}$ (a contradiction).
\end{proofEnd}

Our mechanism in Section \ref{sec:optimize} approximates Problem \ref{problem1} with an optimization problem that has fewer constraints and avoids discrete optimization when breaking ties.

\section{Related Work}\label{sec:related}

Developing differentially private algorithms under accuracy constraints is an underdeveloped area of research. Wu et al. \cite{accuracyfirst} considered the problem of setting privacy parameters to achieve the desired level of accuracy in a certain machine learning task like logistic regression. Our work focuses on linear queries and has multiple (not just one) accuracy constraints. We consider this to be the difference between fitness-for-use (optimizing to support multiple applications within a desired accuracy level) vs. accuracy constrained differential privacy (optimizing for a single overall measure of quality). 

The Matrix Mechanism \cite{LMHMR15,mckenna2018optimizing,YYZH16} answers linear queries while trying to minimize the sum squared error of the queries (rather than per-query fitness for use error). Instead of answering the workload matrix directly, it solves an optimization problem to find a different set of queries, called the strategy matrix. It adds independent noise to query answers obtained from the strategy matrix and then recovers the answer to workload queries from the noisy strategy queries. The major challenge of Matrix Mechanism is that the optimization problem is hard to solve. Multiple mechanisms \cite{YZWXYH12, li2012adaptive, yuan2015optimizing, mckenna2018optimizing} have been proposed to reformulate or approximate the Matrix Mechanism optimization problem. For pure differential privacy (i.e., $\delta=0$),  the solution is often sub-optimal  because of non-convexity. Yuan et al. \cite{YYZH16} propose a convex problem formulation for $(\epsilon,\delta)$-differential privacy and provided the first known optimal instance of the matrix mechanism. Although their work solves the total error minimization problem, the strategy may fail to satisfy accuracy constraints for \emph{every} query.
Prior to the matrix mechanism, the search for query strategies was done by hand, often using hierarchical/lattice-based queries (e.g., \cite{privelet,hay2009boosting,hbtree,dingcube}) and later by selecting an optimal strategy from a limited set of choices (e.g., \cite{yaroslavtsev2013accurate,hbtree,li2012adaptive}).
The Matrix-Variate Gaussian (MVG) Mechanism \cite{chanyaswad2018mvg} is an extension in a different direction that is used to answer matrix-valued queries, such as computing the covariance matrix of the data. It does not perform optimization to decide how to best answer a workload query.

Work related to  the factorization mechanism \cite{edmonds2020power,nikolov2014new}, is most closely related to ours. Starting from a different error metric, they arrived at a similar optimization to Problem \ref{problem2}. They focus on theoretical optimality properties, while we focus on special-purpose algorithms for the optimziation problems, hence approximately optimizing our fitness-for-use and their joint error criteria.

The Laplace and Gaussian mechanisms \cite{dwork06Calibrating, diffpbook} are the most common building blocks for differential privacy algorithms, adding noise from the Laplace or Gaussian noise to provide $(\epsilon,0)$-differential privacy and $(\epsilon,\delta)$-differential privacy, respectively. Other distributions are also possible (e.g., \cite{geng2015optimal, geng2020tight, liu2018generalized,ghosuniversal,discgauss}) and their usage depends on application requirements (specific privacy definition and measure of error).

Our work and the matrix/factorization mechanism work are examples of data-independent mechanisms -- the queries and noise structure does not depend on the data. There are many other works that focus on data-dependent mechanisms \cite{cormode2012differentially,li2014data,kellaris2013practical,qardaji2013differentially, kotsogiannis2017pythia,mwem}, where the queries receiving the noise depend on the data. These mechanisms reserve some privacy budget for estimating properties of the data that help choose which queries to ask. For example, the DAWA Algorithm \cite{li2014data} first privately learns a partitioning of the domain into buckets that suits the input data well and then privately estimates counts for each bucket. While these algorithm may perform well on certain dataset, they can often be outperformed on other datasets by data-independent mechanisms \cite{dpcomp}. A significant disadvantage of data-dependent algorithms is that they cannot provide closed-form  error estimates for query answers (in many cases, they cannot provide any accurate error estimates). Furthermore, data-dependent methods also often produce \emph{biased} query answers, which can be undesirable for subsequent analysis, as discussed in Section \ref{subsec:desiderata}.


\section{Theoretical Analysis\label{sec:theo}}\label{sec:theory}
In this section, we theoretically analyze the solution to Problem \ref{problem1}. We prove uniqueness results for the solution and derive results that simplify the algorithm construction (for Section \ref{sec:optimize}). \addcolor{We first show that optimizing per-query error targets is a fundamentally different problem than optimizing for total squared error, as was done in prior work (e.g., \cite{YYZH16,LMHMR15,YZWXYH12,hbtree}). Our results show that there are natural problems where algorithms that optimize for sum of squared errors can have maximum per-query errors that are $O(\sqrt{d})$ times larger than optimal ($d$ is the domain size).}


\addcolor{\subsection{Analytical Case Study}\label{sec:toy}
Suppose the dataset is represented by an $d$-dimensional vector $\mathbf{x}=[x_1,\dots, x_d]^T$. Let the workload matrix consists of identity queries (i.e., for each $i$, what is the value of $x_i$) and the total sum query. Its matrix representation is:
$$\mat{W}=\left[
\begin{matrix}
1 & 0 & 0 & \dots & 0\\
0 & 1 & 0 & \dots & 0\\
\vdots & \vdots & \vdots & \vdots & \vdots\\
0 & 0 & 0 & \dots & 1\\
1 & 1 & 1 & \dots & 1
\end{matrix}
\right]$$
We compare \emph{closed-form} solutions for sum-squared-error and fitness-for-use optimizations for the case where all per-query variance targets are set to $\gamma$. Solutions to both problems can be interpreted as adding \emph{correlated} noise to $\vec{x}$ as follows: $\vec{z}=\vec{x}+N(\vec{0},\mat{\Sigma})$ and then releasing $\mat{W}\vec{z}$.
 By convexity of the sum-squared-error \cite{YYZH16} and fitness-for-use (Section \ref{sec:properties}) problems, and since all  $x[i]$ are treated symmetrically by $\mat{W}$,
the covariance matrices for both problems  should treat the domain elements symmetrically. That is, the correlation between the noise added to $x[i]$ and $x[j]$ (i.e., $\mat{\Sigma}[i,j]$) should be the same as the correlation between the noise added to $x[i^\prime]$ and $x[j^\prime]$. Thus $\mat{\Sigma}$ (and consequently $\mat{\Sigma}^{-1}$) should have the form: 
\begin{align*}
    \mat{\Sigma} &= a  \mat{I} + b  \vec{1}\vec{1}^T & \mat{\Sigma}^{-1} &= \frac{1}{a}\mat{I} - \frac{b}{a^2 + dab}\vec{1}\vec{1}^T
\end{align*}
where a and b are scalars and $\vec{1}$ is the column vector of ones. The eigenvalues are $a+bd$ (with eigenvector $\vec{1}$) and $a$ (for all vectors orthogonal to the vector $\vec{1}$). Hence positive definiteness requires $a>0$ and $a+bd>0$.
The variance for each of the first $d$ queries is $a+b$. For the sum query it is $\mathbf{1}^T \Sigma  \mathbf{1} = d a + d^2 b $. The squared privacy cost is $ \Sigma^{-1}_{i, i} = \frac{a+(d-1)b}{a^2 +d ab}$.

\subsubsection{Sum-squared-error Optimization}

Thus, the optimization problem for sum squared error given a privacy cost $\beta$ is:
\begin{align}
\label{eqn:sum}
    \arg\min_{a, b}  \quad & d (a+b) + (d a + d^2 b )\\
    \nonumber
    s.t. \quad & \frac{a+(d-1)b}{a^2 +d ab} = \beta^2 
    \;\text{ and } \;  a > 0 \;\text{ and }\; a +  b d  > 0
\end{align}


\begin{theoremEnd}[category=section5]{theorem}\label{thm:eqn:sum} For $d\geq 5$, the solution to  Equation \ref{eqn:sum} is: 
\begin{align*}
    a &=  \frac{-3+\mathrm{d}}{(-1+\mathrm{d}) -\sqrt{1+\mathrm{d}}} \frac{1}{\beta^{2}}\\
    b &=  \frac{(-3+\mathrm{d})\left(2 -\sqrt{1+\mathrm{d} }\right)}{\left((-1+\mathrm{d}) -\sqrt{1+\mathrm{d} }\right)\left(-1-\mathrm{d} +(-1+\mathrm{d}) \sqrt{1+\mathrm{d} }\right)} \frac{1}{\beta^2}\\
\end{align*}
\end{theoremEnd}
\begin{proofEnd}
This result  follows from Theorem \ref{thm:eqn:sum:strong} by setting $w=1$. In particular, plugging in $w=1$ into the statement of that theorem, we get
\begin{align*}
    a &=\frac{\left(-3+d\right)}{(-1+d)  - \sqrt{\left(d+1\right) }} \frac{1}{\beta^{2}}\\
 b 
 &= \frac{\left(2- \sqrt{d+1}\right)\left(-3+d\right)}{\left((-1+d) -\sqrt{d+1}\right)\left(-d -1 +(-1+d)\sqrt{d+1}\right)} \frac{1}{\beta^{2}}
\end{align*}
\end{proofEnd}

\begin{theoremEnd}[category=section5, all end]{theorem}\label{thm:eqn:sum:strong}
When $d\geq 5$, $w^2\geq 1$, the solution to the optimization problem
\begin{align}
    \arg\min_{a, b}  \quad & d(a+b) + \frac{1}{w^2}(d a + d^2 b )\\
    \nonumber
    s.t. \quad & \frac{a+(d-1)b}{a^2 +d ab} = \beta^2 \\
        \nonumber
    and \quad & a > 0, a+ b d > 0
\end{align}
 is 
\begin{align*}
    a &= \frac{-1+(-2+d)w^2}{(-1+d)w^2 - w\sqrt{d+w^2}} \frac{1}{\beta^2}  \\
    b &= \left(\frac{1 + w^2 - w\sqrt{d+w^2}}{(-1+d)w^2 - w\sqrt{d+w^2}}\right) \left(\frac{-1+(-2+d)w^2} {-d -w^2  +  (d-1)w\sqrt{d+w^2}} \right) \frac{1}{\beta^2} \\
\end{align*}

\end{theoremEnd}
\begin{proofEnd}\phantom{.}

\textbf{Step 1: rewrite the optimization problem.}

Note that
\begin{align*}
    \Sigma ^{-1} \equiv (a\mat{I} + b\vec{1}\vec{1}^T)^{-1} = \frac{1}{a}\mathbf{I} - \frac{b}{a^2+d ab}  \mathbf{1}\mathbf{1}^T 
\end{align*}
Let us define $x = \frac{1}{a}$ and $y = -\frac{b}{a^2+dab}$. To recover $a$ and $b$ from $x$ and $y$, we note that
\begin{align*}
 \Sigma \equiv  (x\mat{I} + y \vec{1}\vec{1}^
    T)^{-1} = \frac{1}{x}\mat{I} - \frac{y}{x^2+d x y}  \vec{1}\vec{1}^T 
\end{align*}
so that 
\begin{align*}
    a &= 1/x\\
    b &= - \frac{y}{x^2+d x y} 
\end{align*}
Substituting $x$ and $y$ into the objective function, we get
\begin{align*}
    \frac{d}{x} - \frac{d y}{x^2 + d x y} + \frac{1}{w^2} (\frac{d}{x} - \frac{d^2 y}{x^2 + d x y}) 
    & = \frac{d(1+\frac{1}{w^2}) x + d(d-1)y}{x^2+d x y} \\
\end{align*}

Meanwhile, the first constraint of the optimization problem is the same as 
\begin{align*}
    x+y=\beta^2
\end{align*}

Recalling that the eigenvalues of  $a\mat{I} + b\vec{1}\vec{1}^T$
 are $a+bd$ (with eigenvector $\vec{1}$) and $a$ (for all vectors orthogonal to the vector $\vec{1}$). Then this matrix is positive definite if and only if $a>0$ and $a+bd>0$. Also this matrix is positive definite if and only if its inverse, which is $x\mat{I} + y\vec{1}\vec{1}^T$, is positive definite, which means that the pair of constraints $a>0$ and $a+bd>0$ are equivalent to $x>0$ and $x+yd>0$.

Hence the optimziation problem can be rewritten as
\begin{align}
    \arg\min_{x, y} \quad &\frac{d(1+\frac{1}{w^2}) x + d(d-1)y}{x^2+d x y} \\
    \nonumber
    s.t. \quad &  x + y = \beta^2 \\
        \nonumber
    and \quad & x > 0, x + y d > 0
\end{align}

We can plug in $y= \beta^2 - x$ to get the objective function
\begin{align*}
    &\frac{\mathrm{d}\left(1+\frac{1}{\mathrm{w}^{2}}\right) \mathrm{x}+(-1+\mathrm{d}) \mathrm{d}\left(-\mathrm{x}+\beta^{2}\right)}{\mathrm{x}^{2}+\mathrm{d} x\left(-\mathrm{x}+\beta^{2}\right)} \\
    &=\frac{d(w^2 + 1)x + d (1-d)w^2 x + (-1+d)d w^2 \beta^2}{w^2 x^2 -d w^2 x^2 + d x  w^2 \beta^2 } \\
    &=\frac{d((2-d)w^2+1)x + (-1+d)d w^2 \beta^2}{(1-d)w^2 x^2 + d w^2 \beta^2 x}
\end{align*}
So the optimization problem reduces to
\begin{align*}
\arg\min_x &\frac{d((2-d)w^2+1)x + (-1+d)d w^2 \beta^2}{(1-d)w^2 x^2 + d w^2 \beta^2 x} \\
\text{s.t. } & x >0 \;\text{ and }\; x < \frac{d \beta^2}{d-1}
\end{align*}

\textbf{Step 2: identify when the derivative of the objective function is positive and negative.}

Let $A=(1-d)w^2$, $B=d w^2 \beta^2$, $C = d ((2-d) w^2 +1)$, $D = (-1+d)d w^2 \beta^2$, then the objective function is $f = \frac{C x +D}{A x^2 + B x}$.
\begin{align*}
    \frac{\partial f}{ \partial x} &= \frac{C(A x^2 + B x) - (2 A x + B) (C x + D) }{(A x^2 + B x)^2} \\
    & = \frac{A C x^2 + B C x - (2 A C x^2 + B C x + 2 A D x + B D) }{(A x^2 + B x)^2} \\
    & = \frac{-A C x^2 -2 A D x - B D}{(A x^2 + B x)^2}
\end{align*}
Then we calculate the coefficients,
\begin{align*}
    AC &= (-1+d)d((-1+(-2+d)w^2)w^2 \\
    AD &= (-1+d)d(1-d)w^4 \beta^2 \\
    BD &= (-1+d)d^2 w^4 \beta^4
\end{align*}
Finally we get the derivative of the objective function as
\begin{align*}
    \frac{\partial f}{ \partial x} =-\frac{(-1+d) d\Big(\left(-1+(-2+d) w^{2}\right) x^{2} + 2(1-d) w^{2} \beta^{2} x +d w^{2} \beta^{4}\Big)}{w^{2} \left((1-d) x^2 + d \beta^{2} x \right)^{2} }
\end{align*}
It equals 0 only when the quantity inside the large parentheses is 0, which is a quadratic equation.
Let $A_1 = -1 + (-2 + d) w^2$, $B_1 = 2(1-d) w^2 \beta^2$, $C_1 = d w^2 \beta^4$, the roots of  the quadratic equation
\begin{align*}
    A_1 x^2 + B_1 x + C_1 = 0
\end{align*}
the solutions are known as
\begin{align*}
    x_1 & = \frac{-B_1 - \sqrt{B_1^2 - 4 A_1 C_1}}{2 A_1} \\
    x_2 & = \frac{-B_1 + \sqrt{B_1^2 - 4 A_1 C_1}}{2 A_1}
\end{align*}
Here 
\begin{align*}
    B_1^2 - 4 A_1 C_1 & = 4(1-d)^2 w^4 \beta^4 - 4 d w^2 \beta^4 (-1+(-2+d)w^2) \\
    &= 4 w^2 \beta^4 (w^2 (1-d)^2 - (-d + (-2d+ d^2)w^2)) \\
    &= 4 w^2 \beta^4 (d + w^2)
\end{align*}
So the roots are 
\begin{align*}
    x_1 = \frac{(-1+d)w^2 - w\sqrt{d+w^2}}{-1+(-2+d)w^2}  \beta^2 \\
    x_2 = \frac{(-1+d)w^2 + w\sqrt{d+w^2}}{-1+(-2+d)w^2}  \beta^2
\end{align*}
Since, by assumption,  $d \geq 5, w 
\geq 1$, then the quantity $A_1 \equiv -1+(-2+d)w^2 \geq -3 + d > 0$ and $d(-1+d) > 0$. Noting that the numerator of $ \frac{\partial f}{\partial x}$ is $-(-1+d)d (A_1 x^2 + B_1 x + C_1)$, the last 2 facts imply that the coefficient of $x^2$ in the numerator is negative. Also the denominator of $ \frac{\partial f}{\partial x}$ is a squared term which is always greater than 0. So $ \frac{\partial f}{\partial x} <0$ when $x< x_1$ or $x>x_2$ (since this is where the quadratic in the numerator is negative). $ \frac{\partial f}{\partial x} >0$ when $x_1 < x < x_2$. 

\textbf{Step 3: analyze when the constraints in the optimization problem are satisfied.}

Recall that we also have two constraints in our optimization problem: $0<x< \frac{d\beta^2}{d-1}$.

Next we claim that $0< x_1 < \frac{d\beta^2}{d-1} < x_2$ when $d\geq 5$ and $w^2 \geq 1$. 
Note that the denominator of $x_1$ is $-1+(-2+d)w^2 >0$,
\begin{align*}
    x_1 > 0 \Longleftrightarrow &(-1+d)w^2 - w\sqrt{d+w^2} >0 \\
    \Longleftrightarrow & (d^2 - 2d + 1)w^2 > d+ w^2 \\
    \Longleftrightarrow & w^2(d^2 - 2d) > d 
\end{align*}
Because $d\geq 5$, we know that $d^2-2d >d$. With $w^2 \geq 1$,  $w^2(d^2-2d) \geq d^2-2d > d$ so the last inequality holds.
\begin{align*}
    x_1 <  \frac{d\beta^2}{d-1} \Longleftrightarrow &\frac{(-1+d)w^2 - w \sqrt{d+w^2}}{-1+(-2+d)w^2}  \beta^2 < \frac{d\beta^2}{d-1} \\
    \Longleftrightarrow & w^2 (d^2 - 2d + 1) - w\sqrt{d+w^2} (d-1) < -d + (-2d + d^2) w^2 \\
    \Longleftrightarrow & w^2 + d < w\sqrt{d+w^2} (d-1) \\    
    \Longleftrightarrow & (w^2 + d)^2 < w^2(d+w^2) (d-1)^2 \\
    \Longleftrightarrow & w^2 + d < w^2(d^2 - 2d +1) \\
    \Longleftrightarrow & w^2(d^2 - 2d) > d 
\end{align*}
The last inequality holds when $d\geq 5, w\geq 1$. Next, note that:
\begin{align*}
    x_2 > \frac{d\beta^2}{d-1} \Longleftrightarrow &\frac{(-1+d)w^2 + w\sqrt{d+w^2}}{-1+(-2+d)w^2} \beta^2 > \frac{d\beta^2}{d-1} \\
    \Longleftrightarrow & w^2 (d^2 - 2d + 1) + w\sqrt{d+w^2} (d-1) > -d + (-2d + d^2) w^2 \\
    \Longleftrightarrow & w^2 + d + w\sqrt{d+w^2} (d-1) >0
\end{align*}
The last inequality holds because $d-1>0$, $w^2 +d >0$, $w>0$.

From the above analysis, we know that the objective function $f$ to our rewritten optimization problem  decreases when $x$ increases in $(0, x_1]$, then the objective increases over $(x_1, \frac{d \beta^2}{d-1})$ until it hits the constraint. So the minimal $f$ on $0<x<\frac{d \beta^2}{d-1}$ is obtained at the root $x=x_1$. Using this solution to our transformed objective function, we recover the solution to the original problem as follows (see the relation beween $a, b$ and $x,y$ is step 1 of the proof):  
\begin{align*}
    a &= \frac{1}{x} 
     = \frac{-1+(-2+d)w^2}{(-1+d)w^2 - w\sqrt{d+w^2}} \frac{1}{\beta^2}
\end{align*}
\begin{align*}
    y &= \beta^2 - x \\
    &= \frac{-1+(-2+d)w^2 - (-1+d)w^2 + w\sqrt{d+w^2}}{-1+(-2+d)w^2}\beta^2 \\
    &=  \frac{-1 - w^2 + w\sqrt{d+w^2}}{-1+(-2+d)w^2}\beta^2
\end{align*}
\begin{align*}
    y/x 
    &= \frac{-1 - w^2 + w\sqrt{d+w^2}}{(-1+d)w^2 - w\sqrt{d+w^2}} 
\end{align*}
\begin{align*}
    x + d y &= \frac{(-1+d)w^2 - w\sqrt{d+w^2} -d - d w^2 + d w\sqrt{d+w^2}}{-1+(-2+d)w^2} \beta^2 \\
    &= \frac{-d -w^2  +  (d-1)w\sqrt{d+w^2}}{-1+(-2+d)w^2}  \beta^2
\end{align*}
\begin{align*}
    b &= -\frac{y}{x^2+d x y} = -\frac{y/x}{x+d y}   \\
    &= - \left(\frac{-1 - w^2 + w\sqrt{d+w^2}}{(-1+d)w^2 - w\sqrt{d+w^2}}\right) \left(\frac{-1+(-2+d)w^2} {-d -w^2  +  (d-1)w\sqrt{d+w^2}} \right) \frac{1}{\beta^2} \\
    &= \left(\frac{1 + w^2 - w\sqrt{d+w^2}}{(-1+d)w^2 - w\sqrt{d+w^2}}\right) \left(\frac{-1+(-2+d)w^2} {-d -w^2  +  (d-1)w\sqrt{d+w^2}} \right) \frac{1}{\beta^2}
\end{align*}
\end{proofEnd}

\subsubsection{Fitness-for-use Optimization}
Fitness-for-use optimization with target variance $\gamma$ for all queries can be written as: 
\begin{align}
\label{eqn:fit}
    \arg\min_{a, b} \quad & \frac{a+(d-1)b}{a^2 +d ab}\\
    s.t.  \quad & a+b\leq \gamma \; \text{ and }\; a d+b d^2 \leq \gamma \; \text{ and }\; a>0   \; \text{ and }\; a+b d>0\nonumber
\end{align}

\begin{theoremEnd}[category=section5]{theorem}\label{thm:eqn:fit} When $d\geq 5$, the solution to Equation \ref{eqn:fit} is  $a = (\frac{d+1}{d}) \gamma $, $b = -(\frac{1}{d}) \gamma$, with squared privacy cost $ \frac{2 d}{1+d}\frac{1}{\gamma} $.
\end{theoremEnd}

\begin{proofEnd}
We plug in $k=1$ into the result of Theorem \ref{thm:eqn:fit:strong}.
\begin{align*}
    (\text{privacy cost})^2 &= \frac{2d - d^2 - d^2}{ (1 - d^2)}\frac{1}{\gamma} =\frac{2d(1-d)}{(1-d)(1+d)}\frac{1}{\gamma}=\frac{2d}{1+d}\frac{1}{\gamma}\\
    a &= \left(\frac{-1+d^2}{(-1+d)d}\right) \gamma =\left(\frac{(d-1)(d+1)}{d(d-1)}\right)\gamma = \left(\frac{d+1}{d}\right)\gamma\\
    b &= \left(\frac{1-d}{(-1+d)d}\right) \gamma = -\left(\frac{1}{d}\right)\gamma
\end{align*}
\end{proofEnd}

\begin{theoremEnd}[category=section5, all end]{theorem}\label{thm:eqn:fit:strong}
When $d\geq 5$ and $0 < k < d$, $\gamma > 0$, the solution to 
\begin{align}
    \arg\min_{a, b} \quad & \frac{a+(d-1)b}{a^2 +d ab}\\
    \nonumber
    s.t.  \quad & a+b\leq \gamma \\
    \nonumber
    and \quad & a d+b d^2 \leq k \gamma \\
    \nonumber 
    and \quad & a>0, a+b d>0
\end{align}
is

\begin{align*}
    (\text{privacy cost})^2 &= \frac{2k d - d^2 - k d^2}{k (k - d^2)}\frac{1}{\gamma} \\
    a &= \left(\frac{-k+d^2}{(-1+d)d}\right) \gamma \\
    b &= \left(\frac{k-d}{(-1+d)d}\right) \gamma
\end{align*}
\end{theoremEnd}
\begin{proofEnd}
The derivative of the objective function with respect to $a$ is
\begin{align*}
    \frac{\partial  }{\partial a}& = \frac{a^2 + d a b - (a+(d-1)b)(2a+db)}{a^{2}(a+b d)^{2}} \\
    & = \frac{a^2 + d a b - (2a^2 + 2(d-1) a b + a d b + (d-1) d b^2)}{a^{2}(a+b d)^{2}} \\
    & = \frac{- a^2  -   2(d-1) a b 
    - (d-1) d b^2}{a^{2}(a+b d)^{2}} \\
    & = \frac{- a^2  -   2(d-1) a b - (d-1)^2 b^2 + (d-1)^2 b^2
    - (d-1) d b^2}{a^{2}(a+b d)^{2}} \\
    & = \frac{- (a + (d-1)b)^2 - (d-1) b^2 }{a^{2}(a+b d)^{2}} < 0
\end{align*}
This means the objective function decreases when $a$ increases. The derivative of the objective function with respect to $b$ is
\begin{align*}
    \frac{\partial  }{\partial b} &=\frac{(d-1)(a^2 + d a b) - (a+(d-1)b)d a}{a^{2}(a+b d)^{2}} \\
    & = \frac{(d-1)a^2 +(d-1)d a b  - d a^2 - (d-1)d a b }{a^{2}(a+b d)^{2}} \\
    & = \frac{- a ^2 }{a^{2}(a+b d)^{2}}  < 0
\end{align*}

This means that the objective function decreases when $b$ increases.
Thus, to minimize the objective function, we need to increase $a$ and $b$ as much as we can  under the constraints
 $a+b \leq \gamma$ and $a d + b d^2 \leq k \gamma$. 
 We will show that under the optimal solution, both constraints have to hold with equality.
\begin{align*}
    a+b &= \gamma \\
    a d + b d^2 &= k \gamma
\end{align*}
To show this, we consider the 3 alternative cases where equality doesn't hold.

\textbf{\underline{Case 1}}: 
\begin{align*}
    a+b &< \gamma \\
    a d + b d^2 &< k \gamma
\end{align*}
If both qualities don't hold, we can always find a larger $a$ or $b$ that still satisfies the constraints (and thus decreases the objective function since it is decreasing in $a$ and $b$). Thus the optimal solution cannot occur under case 1.

\textbf{\underline{Case 2}}:
\begin{align*}
    a+b &= \gamma \\
    a d + b d^2 &< k \gamma
\end{align*}
This means  $b= \gamma - a$ and we plug it   into the objective function to get 
\begin{align*}
    \frac{ a  +(-1+d)(-a+\gamma)}{a^{2}+ a  d(-a+\gamma)} 
    &=\frac{(2-d) a  + (-1+d)\gamma}{(1-d)a^{2}+  d \gamma a  } \\
\end{align*}
Then the optimization problem reduces to 
\begin{align*}
\arg\min_a& \quad  \frac{(2-d) a  + (-1+d)\gamma}{(1-d)a^{2}+  d \gamma a  }  \\
\text{s.t. }& \quad a > \frac{d^2 - k }{d^2-d}\gamma \\
& \quad a >0, a < \frac{d \gamma}{d-1}
\end{align*}

Let $A=1-d$, $B=d \gamma $, $C = 2-d$, $D = (-1+d) \gamma $, then the objective function is $ \frac{C a +D}{A a^2 + B a}$.
\textbf{We then find for which values of $a$ is the objective function increasing/decreasing}.
\begin{align*}
    \frac{\partial }{ \partial a} &= \frac{C(A a^2 + B a) - (2 A a + B) (C a + D) }{(A a^2 + B a)^2} \\
    & = \frac{A C a^2 + B C a - (2 A C a^2 + B C a + 2 A D a + B D) }{(A a^2 + B a)^2} \\
    & = \frac{-A C a^2 -2 A D a - B D}{(A a^2 + B a)^2}
\end{align*}
Then we calculate the coefficients,
\begin{align*}
    AC &= (1-d)(2-d) \\
    AD &= (1-d)(-1+d)\gamma \\
    BD &= d(-1+d) \gamma^2
\end{align*}
Finally we get the derivative of the objective function as
\begin{align*}
    \frac{\partial }{ \partial a} =-\frac{(-1+d) \Big(\left(-2+d \right) a^{2} + 2(1-d)\gamma  a +d \gamma^2\Big)}{ \left((1-d) a^2 + d \gamma a \right)^{2} }
\end{align*}
The quadratic inside the large parentheses determines the sign of the derivative, since the terms that multiply it are positive for $d\geq 5$ and the negative sign in front means the derivative is positive when the quadratic is negative.
Let $A_1 = -2 + d $, $B_1 = 2(1-d)\gamma$, $C_1 = d \gamma^2$. Then this quadratic is equal to 
\begin{align*}
    A_1 a^2 + B_1 a + C_1 = 0
\end{align*}
the solutions are known as
\begin{align*}
    a_1 & = \frac{-B_1 - \sqrt{B_1^2 - 4 A_1 C_1}}{2 A_1} \\
    a_2 & = \frac{-B_1 + \sqrt{B_1^2 - 4 A_1 C_1}}{2 A_1}
\end{align*}
Here 
\begin{align*}
    B_1^2 - 4 A_1 C_1 & = 4(1-d)^2 \gamma^2 - 4 d \gamma^2 (-2+d) \\
    &= 4 \gamma^2 (1 - 2d + d^2  +2d - d^2) \\
    &= 4 \gamma^2
\end{align*}
So the roots are 
\begin{align*}
    a_1 &= \frac{-(1-d) - 1}{-2+d}  \gamma =  \gamma\\
    a_2 &= \frac{-(1-d) + 1}{-2+d}  \gamma = \frac{d}{-2 + d} \gamma
\end{align*}

\textbf{Now we analyze the signs of the quadratic for different values of $a$.}
Since, by assumption,  $d \geq 5$, then the quantity $A_1 \equiv -2+d > 0$ and $-1+d > 0$. Noting that the numerator of the derivative of the objective is $-(-1+d) (A_1 a^2 + B_1 a + C_1)$ 
the derivative is negative when $a< a_1$ or $a>a_2$ (since these are the roots of the quadratic). The derivative is positive (hence the objective function is increasing) when $a_1 < a < a_2$. 

Recall that we also have three constraints in our optimization problem: $a > \frac{d^2-k}{d^2-d} \gamma$, $ a > 0$ and $a < \frac{d \gamma}{d-1}$.

Next we claim that under our assumptions $d\geq 5$, $0 < k < d$, $\gamma >0$, we have $0 < a_1 < \frac{d^2-k}{d^2-d}\gamma < \frac{d}{d-1}\gamma < a_2$.

From the value of $a_1 = \gamma$ we know that,
\begin{align*}
    0 < a_1  & \Longleftrightarrow 0 < \gamma 
\end{align*}
Because $d\geq 5$, we know that $d^2 - d > 0$, so
\begin{align*}
     a_1 <  \frac{d^2-k}{d^2-d}\gamma  & \Longleftrightarrow \gamma  < \frac{d^2-k}{d^2-d}\gamma  \\
      & \Longleftrightarrow d^2 - d< d^2 -k \\
       & \Longleftrightarrow k < d
\end{align*}
Because $d\geq 5$, we have $d > 0$, $d> 1$, so
\begin{align*}
       \frac{d^2-k}{d^2-d}\gamma < \frac{d}{d-1}\gamma 
      & \Longleftrightarrow d^2 - k< d^2 \\
       & \Longleftrightarrow k > 0
\end{align*}
Because $d \geq 5$, we have $d>2$, so
\begin{align*}
    \frac{d}{d-1}\gamma < a_2  & \Longleftrightarrow \frac{d}{d-1}\gamma < \frac{d}{-2+d}\gamma  \\
    & \Longleftrightarrow -2+d < -1+d \\
    & \Longleftrightarrow -2 < -1
\end{align*}

From the above analysis we can see that the feasible set for $a$ is $(\frac{d^2-k}{d^2-d}\gamma, \frac{d}{d-1}\gamma)$ and the objective function is increasing in $a$ on this set (because its derivative is positive). 
This means that when $a$ decreases, the privacy cost decreases. If $\frac{d^2 - k }{d^2-d}\gamma < a < \frac{d}{d-1}\gamma$, we can always find a smaller privacy cost by decreasing $a$, meaning that the strict inequalities of Case 2 will not result in an optimal solution (since the lower bound $\frac{d^2 - k }{d^2-d}\gamma < a$ is the result of taking the strict inequality $ad+bd^2<k\gamma$ assumed by Case 2 and substituting the equation for $b$, as done at the beginning of Case 2). Hence, Case 2 does not contain the optimal solution. 

\textbf{\underline{Case 3}}:
\begin{align*}
    a+b &< \gamma \\
    a d + b d^2 &= k \gamma
\end{align*}
This means $a+bd=k\gamma/d>0$ and so the constraint $a+b d>0$ is automatically satisfied.
This also means $b=\frac{ k \gamma - a d }{d^2}$ and we can plug it into the objective function to get 
\begin{align*}
     \frac{a + (d-1) \frac{ k \gamma - a d }{d^2}}{ a^2 + d a \frac{ k \gamma - a d }{d^2}} 
    &= \frac{a d^2 + d k \gamma - a d^2 - k \gamma + a d}{a^2d^2 + a d k \gamma - a^2 d^2} \\
    & = \frac{(-1+d)k \gamma + a d}{a d k \gamma} \\
    & = \frac{-1 + d}{a d} + \frac{1}{k \gamma}
\end{align*}

Then the optimization problem reduces to
\begin{align*}
\arg\min_a&=\frac{-1 + d}{ d} \frac{1}{a} + \frac{1}{k \gamma} \\
\text{s.t. }&a < \frac{d^2 \gamma - k \gamma}{d^2 - d} \quad\text{(i.e., substitute expression for $b$ in $a+b< \gamma$)}\\
&a >0 
\end{align*}
When $d \geq 5$, clearly the objective function decreases as $a$ increases. So we can keep decreasing the objective function by setting $a=\frac{d^2 \gamma - k \gamma}{d^2 - d}$ so that the upper bound constraint is turned into an equality, which means $a+b=\gamma$, which means that the optimal solution cannot be in Case 3.

\vspace{0.3cm}
Thus, after examining these three cases, we see that equality has to hold (i.e., $a+b=\gamma$ and $ad+bd^2=k\gamma$)
and this means that 
\begin{align*}
    a &= \left(\frac{-k+d^2}{(-1+d)d}\right) \gamma \\
    b &= \left(\frac{k-d}{(-1+d)d}\right) \gamma
\end{align*}
One can easily verify that when $d \geq 5, 0<k<d$, we have $a>0, a+b d>0$.

Noting that the objective function of the original problem is the privacy cost squared, which is $\frac{a + (d-1)b}{a^2 + dab}= \frac{1}{a} - \frac{b}{a^2+dab}$. Hence
\begin{align*}
    (\text{privacy cost})^2 &=\frac{1}{a} - \frac{b}{a^2+dab}
    = \frac{1}{a}(1 -\frac{b}{a+db})\\ 
    &=\frac{1}{\gamma}\frac{d(d-1)}{d^2-k}\left(1-\frac{k-d}{(d^2-k) + d(k-d)}\right)\\
    &=\frac{1}{\gamma}\frac{d(d-1)}{d^2-k}\left(1-\frac{k-d}{dk-k}\right)\\
        &=\frac{1}{\gamma}\frac{d(d-1)}{d^2-k}\frac{dk-2k+d}{k(d-1)}\\
        &=\frac{1}{\gamma}\frac{d}{d^2-k}\frac{dk-2k+d}{k}\\
        &=\frac{1}{\gamma}\frac{d^2k - 2dk + d^2}{k(d^2-k)}
\end{align*}
\end{proofEnd}

\subsubsection{Comparison}
To compare the two mechanisms, we can make their privacy costs equal by   setting $\beta^2 = \frac{2 d}{1+d}\frac{1}{\gamma}$.

\begin{theoremEnd}[category=section5]{theorem}\label{thm:eqn:ratio} When the two mechanisms have the same privacy cost, the maximum ratio of query variance to its desired variance bound $\gamma$ 
is $O(\sqrt{d})$ for sum-squared-error and $1$ for fitness-for-use.
\end{theoremEnd}
\begin{proofEnd}
Using theorem 7,\ref{thm:eqn:fit}, we set the privacy costs to be equal by setting $\beta^2  = \frac{2 d}{1+d}\frac{1}{\gamma}$.
Using theorem \ref{thm:eqn:sum}, we have for sum-squared-error problem,
\begin{align*}
    a+b & = \frac{(-3+d)(-1-d + (-1 + d)\sqrt{1+d} + 2 - \sqrt{1+d})}{(-1+d-\sqrt{1+d})(-1-d +(-1+d)\sqrt{1+d}))} \frac{1}{\beta^2} \\
    &= \frac{(-3+d)(1-d + (-2 +d) \sqrt{1+d})(1+d)}{(-1+d-\sqrt{1+d})(-1-d +(-1+d)\sqrt{1+d}))2 d} \gamma 
\end{align*}
\begin{align*}
    ad+ b d^2 
    &=\frac{(-3+d)(-d-d^2 + (-d + d^2)\sqrt{1+d} + 2d^2 - d^2 \sqrt{1+d})}{(-1+d-\sqrt{1+d})(-1-d +(-1+d)\sqrt{1+d}))} \frac{1}{\beta^2}  \\
    &= \frac{(-3+d)(-d+d^2 -d \sqrt{1+d})}{(-1+d-\sqrt{1+d})(-1-d +(-1+d)\sqrt{1+d}))} \frac{1}{\beta^2} \\
    &= \frac{(-3+d)d}{(-1-d +(-1+d)\sqrt{1+d}))} \frac{1}{\beta^2} \\
    &= \frac{(-3+d)(1+d)}{2(-1-d +(-1+d)\sqrt{1+d}))} \gamma 
\end{align*}

Let us consider first the variance ratio for $\frac{a+b}{\gamma}$ (which is the variance of each of the identity queries). The numerator is $\theta(d^3\sqrt{d})$ and the denominator is also $\theta(d^3\sqrt{d})$ so this ratio is $O(1)$.

Now let us consider the variance ratio for the sum query: $\frac{ad+bd^2}{\gamma}$. The numerator is $\theta(d^2)$ and denominator is $\theta(d\sqrt{d})$, so this ratio is $O(\sqrt{d})$.

%
%

Meanwhile the ratio for the fitness-for-use problem is 1, by definition of the optimization problem.
\end{proofEnd}


This result shows that sum-squared-error optimization and fitness-for-use optimization produce very different solutions, hence fitness-for-use optimization is needed for applications that demand per-query accuracy constraints.}

\subsection{Properties of the Problem and Solution}\label{sec:properties}

We next theoretically analyze the problem and its solution space. The first result is convexity of the optimization problem.
\begin{theoremEnd}[category=section5]{theorem}\label{thm:convex} The optimization problem in Equation \ref{eq:problem1}  is convex.
\end{theoremEnd}
\begin{proofEnd}
Note that the objective function ($\alpha$) is linear (convex). The constraints $(\mat{L}\mat{\Sigma} \mat{L}^T)[i,i]\leq c_i$ are linear in $\mat{\Sigma}$ and hence convex. The set of symmetric positive definite matrices is convex. Finally, the function $\vec{b}^T\mat{\Sigma}^{-1}\vec{b}$ is convex over the (convex) space of symmetric positive definite matrices \cite{YYZH16} and so forms a convex constraint.
\end{proofEnd}

Some convex optimization problems have no solutions because they are not \emph{closed}. As a simple example, consider $\arg\min\limits x^3+x$ s.t., $x > 0$. The value $x=0$ is ruled out by the constraint and no positive value of $x$ can be optimal (i.e. dividing a candidate solution by 2 always improves the objective function). A similar concern holds for positive definite matrices -- the set of  positive definite matrices is not closed, but the set of positive semi-definite matrices is closed. The next result shows that even if we allowed $\mat{\Sigma}$ to be positive semi-definite (hence guaranteeing an optimal solution exists), optimal solutions will still be positive definite.

\begin{theoremEnd}[category=section5]{theorem}\label{thm:pd}  If the fitness-for-use variance targets $c_i$ are all positive  then optimization problem in  Equation \ref{eq:problem1} (Problem \ref{problem1}) is feasible and all optimal solutions for $\mat{\Sigma}$ have smallest eigenvalue $\geq \chi$ for some fixed $\chi>0$ (i.e., they are symmetric positive definite). 
\end{theoremEnd}
\begin{proofEnd}
Note again that in our problem,  the $k\times d$ matrix $\mat{B}$ has linearly independent rows, so $k\leq d$. This means that $\mat{B}\mat{B}^T$ is invertible and hence positive definite. Denote its smallest eigenvalue by $\chi_{\mathbf{B}}$ and note that therefore $\chi_{\mat{B}}>0$ 

Next we show the problem constraints are feasible. If one sets $\mat{\Sigma}^\prime = (\min_{i} \frac{c_i}{||\vec{\ell}_i||^2}) \mat{I}$ (where the $\vec{\ell}_i$ are the \emph{rows} of $\mat{L}$), then this $\mat{\Sigma}^\prime$ satisfies all of the constraints and achieves the objective function value of  $ \frac{\max_j ||\vec{b}_j||^2}{\min_i \frac{c_i}{||\vec{\ell}_i||^2}}$ (where the $\vec{b}_j$ are the \emph{columns} of the basis matrix).

Set $$\chi = \frac{\chi_{\mat{B}}}{d}\Big/ \left(\frac{\max_j ||\vec{b}_j||^2}{\min_i \frac{c_i}{||\vec{\ell}_i||^2}}\right)$$

Next, note that for any symmetric positive definite matrix $\mat{\Sigma}$,  a number $v$ is an eigenvalue of $\mat{\Sigma}$ is and only if $1/v$ is an eigenvalue of $\mat{\Sigma^{-1}}$. So for any symmetric positive definite $\mat{\Sigma}$ that satisfies the constraints of Equation \ref{eq:problem1}, let $\sigma$ be its smallest eigenvalue with corresponding eigenvector $\vec{s}$. Then the largest eigenvalue of $\mat{\Sigma}^{-1}$ is $1/\sigma$.
If $\sigma < \chi$, then the value of the objective function for such a $\mat{\Sigma}$ is
\begin{align*}
\max_{i=1,\dots, d} \vec{b}_i^T \mat{\Sigma}^{-1}\vec{b}_i &\geq  \max_{i=1\dots d} \frac{1}{\sigma}(\vec{s}\cdot\vec{b}_i)^2 > \max_{i=1\dots d} \frac{1}{\chi}(\vec{s}\cdot\vec{b}_i)^2 \\
&\geq \frac{1}{d}\sum_{i=1}^d \frac{1}{\chi}(\vec{s}\cdot\vec{b}_i)^2\\
&=\frac{1}{\chi}\frac{1}{d} \vec{s}^T BB^T\vec{s}\geq \frac{\chi_\mat{B}}{\chi}\frac{1}{d}\\
&= \frac{\max_j ||\vec{b}_j||^2}{\min_i \frac{c_i}{||\vec{\ell}_i||^2}}
\end{align*}
so if the smallest eigenvalue of $\mat{\Sigma}<\chi$ then its objective value is strictly worse than that of the feasible value $\mat{\Sigma}^\prime$. Hence all symmetric positive definite optimal solutions have to have eigenvalues that are all larger than $\chi$, and by continuity we cannot have nearly optimal solutions arbitrarily close to the boundary of the set of positive definite matrices (hence the optimal solution is in the interior of the space of positive definite matrices).
\end{proofEnd}

Problem \ref{problem1} asks us to solve the optimization problem in Equation \ref{eq:problem1} and if there are multiple optimal solutions, pick one such $\mat{\Sigma}^\dagger$ that is minimal under the refined privacy ordering $\prec_R$. In principle, since this is a weak ordering (i.e., two distinct mechanisms can have the same exact privacy properties), one can expect there to exist many such minimal solutions with equivalent privacy properties. 

Surprisingly, it turns out that there is a unique solution. The main idea of the proof is to first show that all solutions to Equation \ref{eq:problem1} that are minimal under $\preceq_R$ have the same exact privacy profile (the refined privacy ordering guarantees that the \emph{sorted} privacy profiles are the same, but we show that for solutions of Equation \ref{eq:problem1}, the entire privacy profiles are exactly the same). This places at least $k$ constraints on the covariance matrix $\mat{\Sigma}$ for such solutions. Note that $\Sigma$ has $k(k+1)/2$ parameters in all, so, in general, two matrices with the same privacy profile do not have to be identical. However, our proof then shows that if two matrices have the same privacy profile \emph{and} are solutions to Equation \ref{eq:problem1}, then they must in fact be the same matrix.

\begin{textAtEnd}[category=uniqueness]
We first show that all  solutions to Equation \ref{eq:problem1} that are also minimal under $\preceq_R$ have the same exact privacy profile.
\end{textAtEnd}

\begin{theoremEnd}[category=uniqueness,all end]{theorem}\label{thm:profile}  The diagonal of $B^T\mat{\Sigma}^{-1}B$ is the same for all $\mat{\Sigma}$ that (1) minimize Equation \ref{eq:problem1} and (2) are minimal under the privacy ordering $\preceq_R$ among all solutions to Equation \ref{eq:problem1}. Hence all such minimal $\mat{\Sigma}$ have the same privacy profile.
\end{theoremEnd}

\begin{proofEnd}
Suppose there exist $\mat{\Sigma_1}$ and $\mat{\Sigma_2}$ that both minimize the Equation \ref{eq:problem1} and are minimal under the refined privacy order. By way of contradiction, assume  the diagonals of $\mat{B}^T\mat{\Sigma}_1^{-1}\mat{B}$ are not equal to the diagonals of $\mat{B}^T\mat{\Sigma}_2^{-1}\mat{B}$. Thus there exists at least one column vector $\vec{b}$ of $\mat{B}$ for which $\vec{b}^T\mat{\Sigma}_1^{-1}\vec{b}\neq \vec{b}^T\mat{\Sigma}_2^{-1}\vec{b}$.

Let $\vec{b}_1, \vec{b}_2,\dots,$ be the columns of $\mat{B}$, sorted in decreasing order by $\vec{b}_i^T\Sigma_1^{-1}\vec{b}_i$, with ties broken in decreasing order by $\vec{b}_i^T\Sigma_2^{-1}\vec{b}_i$ (and if that does not resolve ties, then break ties arbitrarily). Choose $i^*$ to be the first $\vec{b}_{i^*}$ (under this ordering) for which $\vec{b}_{i^*}^T\Sigma_1^{-1}\vec{b}_{i^*}\neq \vec{b}_{i^*}^T\Sigma_2^{-1}\vec{b}_{i^*}$. 

This ordering gives us several important facts:

\textbf{Fact 1.} $\vec{b}_{i^*}^T\Sigma_1^{-1}\vec{b}_{i^*}> \vec{b}_{i^*}^T\Sigma_2^{-1}\vec{b}_{i^*}$. To see why this is true, first note that equality is impossible by definition of $i^*$. Second, we note that the statement $\vec{b}_{i^*}^T\Sigma_1^{-1}\vec{b}_{i^*} <  \vec{b}_{i^*}^T\Sigma_2^{-1}\vec{b}_{i^*}$ cannot be true for the following reason. If this "$<$" statement were true, it would mean that $(\mat{\Sigma}_1, \mat{B}) \prec_R (\mat{\Sigma}_2, \mat{B})$
because the $i^*$ \emph{largest} values of  $\profile(\mat{\Sigma}_1,\mat{B})$ are less than or equal to, with at least one strict inequality, a set of $i^*$ values from $\profile(\mat{\Sigma}_2,\mat{B})$). This implies $\mat{\Sigma}_2$ is not minimal under the privacy ordering, which is a contradiction. Thus the only possibility we are left with is  $\vec{b}_{i^*}^T\mat{\Sigma}_1^{-1}\vec{b}_{i^*}> \vec{b}_{i^*}^T\mat{\Sigma}_2^{-1}\vec{b}_{i^*}$.

\textbf{Fact 2.} If $j>i^*$ and $\vec{b}_{i^*}^T\mat{\Sigma}_1^{-1}\vec{b}_{i^*} = \vec{b}_{j}^T\mat{\Sigma}_1^{-1}\vec{b}_{j}$ then $\vec{b}_j^T\mat{\Sigma}_2^{-1}\vec{b}_j < \vec{b}_{i^*}^T\mat{\Sigma}_2^{-1}\vec{b}_{i^*}$. This is a direct consequence of our tie-breaker on the ordering of columns of $\mat{B}$.

\textbf{Fact 3.}  If $j>i^*$ and $\vec{b}_{i^*}^T\mat{\Sigma}_1^{-1}\vec{b}_{i^*} = \vec{b}_{j}^T\mat{\Sigma}_1^{-1}\vec{b}_{j}$ then $\vec{b}_{j}^T\mat{\Sigma}_1^{-1}\vec{b}_{j}> \vec{b}_{j}^T\mat{\Sigma}_2^{-1}\vec{b}_{j}$. This follows directly from Facts 1 and 2.

Consider the covariance matrix $\mat{\Sigma}_t=t\mat{\Sigma}_1 + (1-t)\mat{\Sigma}_2$ for a constant $t\in(0,1)$ that will be chosen later. Since the optimization problem in Equation \ref{eq:problem1} is convex, then $\mat{\Sigma}_t$ is also a minimizer of it (our goal will be to show that for an appropriate choice of $t$, $\mat{\Sigma}_t$ is better than $\mat{\Sigma}_1$ and $\mat{\Sigma}_2$ under the refined ordering $\prec_R$). 
We now claim that there exists choice of $t$ such that the following must hold:
\begin{align}
    \vec{b}_j^T\mat{\Sigma}_t^{-1}\vec{b}_j &\leq \vec{b}_j^T\mat{\Sigma}_1^{-1}\vec{b}_j \quad\text{for $j< i^*$}\label{eq:uniquecase1}\\
    \vec{b}_{i^*}^T\mat{\Sigma}_t^{-1}\vec{b}_{i^*} & < \vec{b}_{i^*}^T\mat{\Sigma}_1^{-1}\vec{b}_{i^*}\label{eq:uniquecase2}\\
    \vec{b}_j^T\mat{\Sigma}_t^{-1}\vec{b}_j & < \vec{b}_{i^*}^T\mat{\Sigma}_1^{-1}\vec{b}_{i^*} \quad\text{for $j > i^*$}\label{eq:uniquecase3}
\end{align}

\textbf{To prove Eq \ref{eq:uniquecase1}},
note that by construction of $i^*$, we have $\vec{b}_j^T\mat{\Sigma}_1^{-1}\vec{b}_j = \vec{b}_j^T\mat{\Sigma}_2^{-1}\vec{b}_j$ for $j<i^*$. Since $\mat{\Sigma}_t$ is a convex combination of $\mat{\Sigma}_1$ and $\mat{\Sigma}_2$, and since the function $f(\mat{\Sigma})=\vec{b}^T\mat{\Sigma}^{-1}\vec{b}$ is convex, Eq \ref{eq:uniquecase1} follows for all $t\in(0,1)$.

\textbf{To prove Eq \ref{eq:uniquecase2}}, note that Fact 1 implies $\vec{b}_{i^*}^T\mat{\Sigma}_1^{-1}\vec{b}_{i^*}> \vec{b}_{i^*}^T\mat{\Sigma}_2^{-1}\vec{b}_{i^*}$ (note the strict inequality). Since the function  $f(\mat{\Sigma})=\vec{b}^T\mat{\Sigma}^{-1}\vec{b}$ is convex and $t\neq 1$ and $t\neq 0$, we have $\vec{b}_{i^*}^T\mat{\Sigma}_t^{-1}\vec{b}_{i^*} \leq t\vec{b}_{i^*}^T\mat{\Sigma}_1^{-1}\vec{b}_{i^*} + (1-t)\vec{b}_{i^*}^T\mat{\Sigma}_2^{-1}\vec{b}_{i^*} < t\vec{b}_{i^*}^T\mat{\Sigma}_1^{-1}\vec{b}_{i^*} + (1-t)\vec{b}_{i^*}^T\mat{\Sigma}_1^{-1}\vec{b}_{i^*} = \vec{b}_{i^*}^T\mat{\Sigma}_1^{-1}\vec{b}_{i^*}$.


\textbf{To Prove Eq \ref{eq:uniquecase3},} Pick any $j>i^*$. We consider the following two subcases: (a) $\vec{b}_j^T\Sigma_1^{-1}\vec{b}_j = \vec{b}_{i*}^T\Sigma_1^{-1}\vec{b}_{i*}$  and (b)  $\vec{b}_j^T\mat{\Sigma}_1^{-1}\vec{b}_j < \vec{b}_{i*}^T\mat{\Sigma}_1^{-1}\vec{b}_{i*}$ (note we cannot have the inequality going the other direction because of the ordering we imposed on the columns of $\mat{B}$).

In subcase (a),  Fact 3 tell us that $\vec{b}_{j}^T\mat{\Sigma}_1^{-1}\vec{b}_{j}> \vec{b}_{j}^T\mat{\Sigma}_2^{-1}\vec{b}_{j}$. 
 Since the function  $f(\mat{\Sigma})=\vec{b}^T\mat{\Sigma}^{-1}\vec{b}$ is convex and $t\neq 1$ and $t\neq 0$, we have $\vec{b}_{j}^T\mat{\Sigma}_t^{-1}\vec{b}_{j} \leq t\vec{b}_{j}^T\mat{\Sigma}_1^{-1}\vec{b}_{j} + (1-t)\vec{b}_{j}^T\mat{\Sigma}_2^{-1}\vec{b}_{j} < t\vec{b}_{j}^T\mat{\Sigma}_1^{-1}\vec{b}_{j} + (1-t)\vec{b}_{j}^T\mat{\Sigma}_1^{-1}\vec{b}_{j} = \vec{b}_{j}^T\mat{\Sigma}_1^{-1}\vec{b}_{j}= \vec{b}_{i*}^T\Sigma_1^{-1}\vec{b}_{i*}$.
Thus Eq \ref{eq:uniquecase3} holds for such $j$ and for all $t\in(0,1)$.

In subcase (b), this is the only situation where $t$ needs to be chosen carefully. Consider all $j$ for which $\vec{b}_j^T\mat{\Sigma}_1^{-1}\vec{b}_j < \vec{b}_{i*}^T\mat{\Sigma}_1^{-1}\vec{b}_{i*}$ and let $\omega_1$ be the largest achievable value of $\vec{b}_j^T\mat{\Sigma}_1^{-1}\vec{b}_j$ among such $j$. Similarly, let $\omega_2$ be the largest achievable value of $\vec{b}_j^T\mat{\Sigma}_2^{-1}\vec{b}_j$ among those $j$. Since $\omega_1 < \vec{b}_{i*}^T\mat{\Sigma}_1^{-1}\vec{b}_{i*}$ and $\omega_2<\infty$, we can find a $t$ (close enough to 1) so that $tc_1 + (1-t)c_2 < \vec{b}_{i*}^T\Sigma_1^{-1}\vec{b}_{i*}$.  For this value of $t$, by convexity, we must have $\vec{b}_j^T\Sigma_t^{-1}\vec{b}_j < \vec{b}_{i*}^T\Sigma_1^{-1}\vec{b}_{i*}$.

Having proven Equations \ref{eq:uniquecase1}, \ref{eq:uniquecase2}, \ref{eq:uniquecase3}, we see that they imply that the $i^*$ largest values of $\vec{b}_j^T\mat{\Sigma}^{-1}_t\vec{b}^j$ are less than or equal to the $i^*$ largest values of  $\vec{b}^T\mat{\Sigma}_1\vec{b}$ with at least one strict inequality. This means that the (descending) sorted privacy profile of $\mat{\Sigma}_t$ is strictly smaller (under lexicographic comparisons) than the sorted privacy profile of $|mat{\Sigma}_1$. Thus $(\mat{\Sigma}_t, \mat{B}) \prec_R (\mat{\Sigma}_1, \mat{B})$, which contradicts the minimality of $\Sigma_1$. Thus the diagonals of $B^T\mat{\Sigma}^{-1}B$ must be the same for all solutions of the optimization problem that are minimal under the refined privacy ordering. 

\end{proofEnd}

\begin{textAtEnd}[category=uniqueness]
Given two positive symmetric definite matrices $\mat{\Sigma}_1$, $\mat{\Sigma}_2$ and a vector $\vec{r}$, the following result considers conditions where $\vec{r}^T \mat{\Sigma}_t^{-1}\vec{r}$ remains the same for every $\mat{\Sigma}_t$ on the line between $\mat{\Sigma}_1$ and $\mat{\Sigma}_2$ and is the second major step needed to prove Theorem \ref{thm:unique}.
\end{textAtEnd}

\begin{theoremEnd}[category=uniqueness,all end]{theorem}\label{thm:invariant2}
Let $\mat{\Sigma}$ be a symmetric positive definite $k\times k$ matrix. Let $\ell\leq k$ be an integer. Let $\mat{R}$ be a matrix.  Suppose there exists a symmetric matrix $\mat{J}$ such that $\mat{\Sigma}+ t\mat{J}$ is invertible for all $t\in [0,1]$.  Then the first $\ell$ diagonals of $\mat{R}^T(\mat{\Sigma}+t\mat{J})^{-1}\mat{R}$ are equal to the first $\ell$ diagonal entries of  $\mat{R}^T\mat{\Sigma}^{-1}\mat{R}$ for all $t\in [0,1]$ if and only if  $(\mat{R}^\prime)^T\mat{\Sigma}^{-1}\mat{J}$ is the 0 matrix, where $\mat{R}^\prime$ consists of the first $\ell$ columns of $\mat{R}$ (i.e., $\mat{R}^\prime$ and $\mat{J}$ are conjugate with respect to $\mat{\Sigma}^{-1}$). Furthermore, if $\ell=k$ and rank$(\mat{R})=k$ then $\mat{J}$ is the 0 matrix. 
\end{theoremEnd}
\begin{proofEnd}
Let $\mat{I}$ denote the identity matrix. Then
\begin{align*}
    0 &= \frac{d \mat{I}}{dt} = \frac{d (\mat{\Sigma}+t\mat{J})(\mat{\Sigma}+t\mat{J})^{-1}}{dt}\\
    &= (\mat{\Sigma}+t\mat{J})\frac{d (\mat{\Sigma}+t\mat{J})^{-1}}{dt} + \frac{d (\mat{\Sigma}+t\mat{J})}{dt}(\mat{\Sigma}+t\mat{J})^{-1}\\
    &=(\mat{\Sigma}+t\mat{J})\frac{d (\mat{\Sigma}+t\mat{J})^{-1}}{dt} + \mat{J}(\mat{\Sigma}+t\mat{J})^{-1} 
\end{align*}
    and so
\begin{align*}
    \frac{d (\mat{\Sigma}+t\mat{J})^{-1}}{dt} &= -(\mat{\Sigma}+t\mat{J})^{-1}J(\mat{\Sigma}+t\mat{J})^{-1}\\
    \frac{d^2 (\mat{\Sigma}+t\mat{J})^{-1}}{dt^2} &= 2(\mat{\Sigma}+t\mat{J})^{-1}J(\mat{\Sigma}+t\mat{J})^{-1}J(\mat{\Sigma}+t\mat{J})^{-1}\\
    \frac{d \mat{R}^T(\mat{\Sigma}+t\mat{J})^{-1}\mat{R} }{dt} &= \mat{R}^T\frac{d (\mat{\Sigma}+t\mat{J})^{-1}}{dt}\mat{R} \\
    &= -\mat{R}^T(\mat{\Sigma}+t\mat{J})^{-1}J(\mat{\Sigma}+t\mat{J})^{-1}\mat{R}\\
    \frac{d^2 \mat{R}^T(\mat{\Sigma}+t\mat{J})^{-1}\mat{R} }{dt^2} &= \\ 2\mat{R}^T(\mat{\Sigma}+t\mat{J})^{-1}&J(\mat{\Sigma}+t\mat{J})^{-1}J(\mat{\Sigma}+t\mat{J})^{-1}\mat{R}
\end{align*}
For the "only if" direction of the theorem,
let $j\leq \ell$ and let $\vec{r}_j$ be the $j^\text{th}$ column of $\mat{R}$. Then we must have $\vec{0}=\frac{d^2 \vec{r}_j^T(\mat{\Sigma}+t\mat{J})^{-1}\vec{r}_j}{dt^2}=\vec{r}_j^T(\mat{\Sigma}+t\mat{J})^{-1}J(\mat{\Sigma}+t\mat{J})^{-1}J(\mat{\Sigma}+t\mat{J})^{-1}\vec{r}_j$ for $t\in[0,1]$.  Evaluating at $t=0$, we see that $\vec{r}_j^T\mat{\Sigma}^{-1}\mat{J}\mat{\Sigma}^{-1}\mat{J}\mat{\Sigma^{-1}}\vec{r}_j=0$. Since $\mat{\Sigma}$ is positive definite, we get $\vec{r}_j^T\mat{\Sigma}^{-1}\mat{J}$ is the $0$ vector so that $\vec{r}_j$ is conjugate to all of the columns of $\mat{J}$.

For the "if" direction, we start with the hypothesis that $(\mat{R}^\prime)^T\mat{\Sigma}^{-1}J$ is the 0 matrix and we will show that the first $\ell$ diagonals of $\mat{R}^(\mat{\Sigma}+t\mat{J})^{-1}\mat{R}$ are equal to the first $\ell$ diagonals of $\mat{R}\mat{\Sigma}^{-1}\mat{R}$.
We note that since $\mat{J}$ is a symmetric $k\times k$ matrix, it can be represented as $\mat{U}\mat{D}\mat{U}^T$, where $\mat{D}$ is a diagonal $k^\prime\times k^\prime$ matrix (for some $k^\prime\leq k$) with no 0s on the diagonal (i.e., the diagonal consists of the nonzero eigenvalues of $\mat{J}$), and $\mat{U}$ has orthonormal columns. Thus $\mat{D}$ is invertible and note that $\mat{\Sigma}$ and $\mat{\Sigma} + t\mat{J}$ are invertible by the hypothesis of this theorem. By the Sherman-Morrison-Woodbury Identity \cite{GoluVanl96},
\begin{align*}
    \left(\mat{\Sigma} + t\mat{J}\right)^{-1} &= \left(\mat{\Sigma} + t \mat{U}\mat{D}\mat{U}^T\right)^{-1}\\
                                &=\mat{\Sigma}^{-1} - \mat{\Sigma}^{-1} \mat{U}\left((t\mat{D})^{-1} +\mat{U}^T\mat{\Sigma}^{-1}\mat{U}\right)^{-1} \mat{U}^T \mat{\Sigma}^{-1} 
\end{align*}
Note that $(\mat{R}^\prime)^T\mat{\Sigma}^{-1}J= (\mat{R}^\prime)^T\mat{\Sigma}^{-1}\mat{U}\mat{D}\mat{U}^T$. Since the rows of $\mat{U}^T$ are linearly independent and $\mat{D}$ is invertible, then $(\mat{R}^\prime)^T\mat{\Sigma}^{-1}\mat{U}\mat{D}\mat{U}^T$ being equal to the 0 matrix implies that  $(\mat{R}^\prime)^T\mat{\Sigma}^{-1}\mat{U}$ is equal to the 0 matrix. Thus for any $j\leq \ell$, the $j^\text{th}$ column vector $\vec{r}_j$ of $\mat{R}^\prime$ has the property that:
\begin{align*}
\lefteqn{    \vec{r}_j^T\left(\mat{\Sigma} + t\mat{J}\right)^{-1} \vec{r}_j}\\
&= \vec{r}_j^T \mat{\Sigma}^{-1}  \vec{r}_j - \vec{r}_j^T \left(\mat{\Sigma}^{-1} \mat{U}\left((t\mat{D})^{-1} +\mat{U}^T\mat{\Sigma}^{-1}\mat{U}\right)^{-1} \mat{U}^T \mat{\Sigma}^{-1}\right) \vec{r}_j\\
     &= \vec{r}_j^T \mat{\Sigma}^{-1}  \vec{r}_j - \left(\vec{r}_j^T \mat{\Sigma}^{-1}\mat{U}\right)\left((t\mat{D})^{-1} +\mat{U}^T\mat{\Sigma}^{-1}\mat{U}\right)^{-1} \mat{U}^T \mat{\Sigma}^{-1}\vec{r}_j\\
     &= \vec{r}_j^T\mat{\Sigma}^{-1}\vec{r}_j
\end{align*}
and so the $j^\text{th}$ diagonal is unchanged, proving the "if" direction.

Finally, if rank$(\mat{R})=k$ then $\mat{R}$ has $k$ linearly independent columns (and also all $k$ of its rows are linearly independent). Without loss of generality, we can assume that these are the first $k$ columns of $\mat{R}$. Let $\mat{R}^\prime$ be the square (invertible) $k\times k$ matrix consisting these $k$ columns. Then $(\mat{R}^\prime)^T\mat{\Sigma}^{-1} \mat{J}=0\Leftrightarrow \mat{\Sigma}^{-1}\mat{J}=0\Leftrightarrow \mat{J}=0$.
\end{proofEnd}

\begin{theoremEnd}[category=uniqueness]{theorem}\label{thm:unique}  In Problem \ref{problem1}, there exists a unique minimal  (under the privacy ordering $\preceq_R$) solution.
\end{theoremEnd}
\begin{proofEnd}
Suppose there exist two solutions $\mat{\Sigma}_1$ and $\mat{\Sigma}_2$ that are minimal under $\preceq_R$. Then by Theorem \ref{thm:profile}, $\vec{b}_i^T\mat{\Sigma}_1^{-1}\vec{b}_i = \vec{b}_i^T\mat{\Sigma}_2^{-1}\vec{b}_i$ for all $i$. 

Next we claim that for any $t\in [0, 1]$, $\mat{\Sigma}_t=t\mat{\Sigma}_1 + (1-t)\mat{\Sigma}_2$ would also be a minimal (under $\preceq_R$) solution. First, by Theorem \ref{thm:convex}, the optimization problem in Equation \ref{eq:problem1} is convex, so $\mat{\Sigma}_t$ is an optimal solution to Equation \ref{eq:problem1}. Furthermore, by convexity of the function  $f(\mat{\Sigma})=\vec{b}^T\mat{\Sigma}^{-1}\vec{b}$, we have $\vec{b}_i^T\mat{\Sigma}_t^{-1}\vec{b}_i \leq t\vec{b}_i^T\mat{\Sigma}_1^{-1}\vec{b}_i + (1-t)\vec{b}_i^T\mat{\Sigma}_2^{-1}\vec{b}_i = \vec{b}_i^T\mat{\Sigma}_1^{-1}\vec{b}_i$ for all $i$ (since $\vec{b}_i^T\mat{\Sigma}_1^{-1}\vec{b}_i = \vec{b}_i^T\mat{\Sigma}_2^{-1}\vec{b}_i$) and therefore $(\mat{\Sigma}_t, \mat{B})\preceq_R \mat{\Sigma}_1, \mat{B})$.

Now apply Theorem \ref{thm:invariant2} by setting $\mat{J}=\mat{\Sigma_1}-\mat{\Sigma}_2$ (noting that $\mat{\Sigma}_2 + t\mat{J}=\Sigma_t$) and $\mat{R}=\mat{B}$. Since $\Sigma_2$ is a $k\times k$ symmetric positive definite matrix and $\mat{B}$ has $k$ rows that are all linearly independent, we get that $\mat{J}=\mat{0}$ so that $\mat{\Sigma}_1=\mat{\Sigma}_2$.
\end{proofEnd}

%
%

\section{Algorithms}\label{sec:optimize}
In this section, we present an algorithm to solve the fitness-for-use problem (Problem \ref{problem1}). Since a major component of Problem \ref{problem1} includes a constrained convex optimization problem, a reasonable choice is to implement an interior point method \cite{NoceWrig06}. However, given the number of constraints involved, we found that such an algorithm would require a significant amount of complexity to ensure scalability, numerical stability, and a feasible initialization for the algorithm to start at. Instead, we closely approximate it with a series of unconstrained optimization problems that are much easier to implement and initialize. For low-dimensional problems, both the interior point method and our approximation returned nearly identical results.

In Section \ref{sec:ineqapp}, we present our reformulation/approximation of the problem. Then, in Section \ref{sec:newton}, we provide a Newton-based algorithm with line search for solving it. We use a trick from the matrix mechanism optimization \cite{LMHMR15} to avoid computing a large Hessian matrix. Then in Section \ref{sec:init} we present initialization strategies that help speed up convergence.

\subsection{Problem Reformulation}\label{sec:ineqapp}
Two of the main features of Problem \ref{problem1} are optimizing the max of the privacy cost while dealing with the variance constraints. Our first step towards converting this to an unconstrained optimization problem is to bring the variance constraints into the objective function. We do this by noting that our variance constraints can be rephrased as a max constraint: 
\begin{align*}
    \forall j, \vec{v}_j \mat{\Sigma} \vec{v}_j^T \leq c_j  \iff \forall j,  \frac{\vec{v}_j \mat{\Sigma} \vec{v}_j^T}{c_j} \leq 1 
    \iff \max_j \frac{\vec{v}_j \mat{\Sigma} \vec{v}_j^T}{c_j} \leq 1
\end{align*}
where $\vec{v}_j$ is the $j^\text{th}$ row of the matrix $\mat{L}$ (recall $\mat{W}=\mat{L}\mat{B}$) and $\vec{c}=[c_1,\dots, c_m]$ are the desired variance upper bounds.
This observation raises the following question: instead of optimizing privacy cost subject to variance constraints, what if we optimize privacy cost \emph{plus} max variance in the objective function (and then rescale $\Sigma$ so that the accuracy constraints are met)? This results in the following problem, which we show is equivalent to Problem \ref{problem1}.

\begin{problem}\label{prob1alt}
Under the same setting as Problem \ref{problem1},  let $\vec{v}_j$ be the $j^\text{th}$ row of the representation matrix $\mat{L}$. Solve the following optimization:
\begin{align}
     \mat{\Sigma} & \leftarrow \arg\min_{\mat{\Sigma}} \max_i\left( \vec{b}_i^T\mat{\Sigma}^{-1}\vec{b}_i\right) + \max_j\left(\frac{\vec{v}_j \mat{\Sigma} \vec{v}_j^T}{c_j}\right) \label{eq:problem1alt}\\
     & \phantom{\leftarrow}\text{s.t. $\mat{\Sigma}$ is symmetric positive definite.}\nonumber
\end{align}
Among all optimal solutions, choose a $\mat{\Sigma}^*$ that is minimal according to the privacy ordering $\prec_R$, and then output $\Sigma^*/\gamma$, where $\gamma = \max_j \frac{\vec{v}_j \mat{\Sigma}^* \vec{v}_j^T}{c_j}$.
\end{problem}
\begin{theoremEnd}[category=approx]{theorem}\label{thm:doublesum}
The optimal solution to Problem \ref{problem1} is the same as the optimal solution to Problem \ref{prob1alt}.
\end{theoremEnd}
\begin{proofEnd}
Consider the $\mat{\Sigma}^*$ produced by solving Problem \ref{prob1alt} and let $\gamma = \max_j \frac{\vec{v}_j \mat{\Sigma}^* \vec{v}_j^T}{c_j}$. 
This means that $\mat{\Sigma}^*$ achieves the variance condition $\vec{v}_j \mat{\Sigma}^* \vec{v}_j^T\leq \gamma c_j$
Now consider solving Problem \ref{problem1} with utility vector $\gamma\vec{c}$. Clearly the optimal solution to that is again $\mat{\Sigma}^*$  -- otherwise we would get a new $\mat{\Sigma}^\prime$ that satisfies the variance constraints and:
\begin{itemize}
    \item $\mat{\Sigma}^\prime$ either has smaller privacy cost, which would mean $\mat{\Sigma}^\prime$ achieves a lower objective function in Equation \ref{eq:problem1alt} (which is a contradiction of the optimality of $\mat{\Sigma}^*$ for Equation \ref{eq:problem1alt}), or
    \item $\mat{\Sigma}^\prime$ has the same privacy cost, but $\mat{\Sigma}^\prime \prec_R \mat{\Sigma}^*$ which is again a contradiction of Problem \ref{prob1alt}.
\end{itemize}
Now, if we take the optimal solution $\mathbf{\Sigma}^*$ to Problem \ref{problem1} with utility vector $\gamma\vec{c}$ and divide it by $\gamma$, the resulting matrix $\mat{\Sigma}^*/\gamma$ matches the original utility constraints: $\vec{v}_j (\mat{\Sigma}^*/\gamma) \vec{v}_j^T \leq c_j$ for all $j$. Using the same reasoning as in Theorem \ref{thm:equiv} (equivalence of Problem \ref{problem1} and Problem \ref{problem2}), $\mat{\Sigma^*}/\gamma$ must be the optimal solution to Problem \ref{problem1} with utility vector $\vec{c}$.
\end{proofEnd}

Next, we can use a common trick \cite{YZWXYH12} for continuous approximation of the max function, known as the \emph{soft max}: $\softmax_t(a_1,.\dots, a_m)= \frac{1}{t}\log(\sum_i e^{t a_i})$, where $t$ is a smoothing parameter. As $t\rightarrow\infty$, the soft max converges to the max function. We can apply this trick to both maxes in Problem \ref{prob1alt}. Now, the nice thing about the soft max function is that  it also approximates the refined privacy ordering. Specifically, if $\mat{\Sigma}_1 \prec_R \mat{\Sigma}_2$, then clearly when $t$ is large enough,
\begin{align*}
        \frac{1}{t}\log(\sum_i \exp(t*\vec{b}_i ^T \mat{\Sigma}_1^{-1} \vec{b}_i)) <     \frac{1}{t}\log(\sum_i \exp(t*\vec{b}_i ^T \mat{\Sigma}_2^{-1} \vec{b}_i)) 
\end{align*}

Plugging  the soft max function in place of the max in Problem \ref{prob1alt}, we finally arrive at our approximation problem:

\begin{problem}\label{problem:app}
Given parameters $t_1$ and $t_2$,
under the same setting as Problem \ref{problem1}, solve the following optimization problem:
\begin{align}
     \mat{\Sigma}^* & \leftarrow \arg\min_{\mat{\Sigma}} \frac{1}{t_1}\log\left(\sum_i \exp(t_1*\vec{b}_i^T\mat{\Sigma}^{-1}\vec{b}_i)\right) \label{eq:probapprox}\\
    &\phantom{\leftarrow \arg\min_{\mat{\Sigma}}} + \frac{1}{t_2}\log\left(\sum_j t_2*\frac{\vec{v}_j \mat{\Sigma} \vec{v}_j^T}{c_j}\right) \nonumber\\
     & \phantom{\leftarrow}\text{s.t. $\mat{\Sigma}$ is symmetric positive definite.}\nonumber
\end{align}
 and then output $\Sigma^*/\gamma$, where $\gamma = \max_j \frac{\vec{v}_j \mat{\Sigma}^* \vec{v}_j^T}{c_j}$.
\end{problem}
Our algorithm solves Equation \ref{eq:probapprox} as a sequence of optimization problems that gradually increase $t_1$ and $t_2$.

\begin{theoremEnd}[category=approx]{theorem}\label{thm:convexapprox}
The optimization Problem \ref{problem:app} is convex.
\end{theoremEnd}
\begin{proofEnd}
The set of symmetric positive definite matrices is convex. For each $i$, the functions $f_i(\mat{\Sigma}) = t_1*\vec{b}_i^T\mat{\Sigma}^{-1}\vec{b}_i$ are convex over the (convex) space of symmetric positive definite matrices \cite{YYZH16}. For each $j$, the functions $g_j(\mat{\Sigma}) = t_2*\frac{\vec{v}_j \mat{\Sigma} \vec{v}_j^T}{c_j}$ is linear in $\mat{\Sigma}$ and hence convex.

According to the composition theorem for convex functions \cite{boyd2004convex}, a composite function $h(f_1(\mat{\Sigma}), \dots, f_k(\mat{\Sigma}))$ is convex if $h$ is convex, $h$ is non-decreasing in each argument, and the $f_i$ are convex. Let $h(z_1,z_2,\dots)=\log\left(\sum_i \exp(z_i)\right)$, this log-sum-exp function is convex and non-decreasing in each argument \cite{boyd2004convex}. Therefore the functions $h(f_1(\mat{\Sigma}), \dots, f_k(\mat{\Sigma}))$ and $h(g_1(\mat{\Sigma}), \dots, g_k(\mat{\Sigma}))$ are both convex, which means the optimization problem in Equation  \ref{eq:probapprox} is convex.
\end{proofEnd}

\addcolor{Although we focus on controlling per-query error, we note that some data publishers may wish to strike a balance between achieving the per-query error targets and minimizing total squared error.
This is easy to achieve by adding a $+\lambda \sum_j \vec{v}_j\mat{\Sigma}\vec{v}^T_j$ term to the objective function of Problem \ref{problem:app}, where $\lambda$ is a weight indicating relative importance of total squared error. This modification changes the objective function gradient by $+\lambda \sum_j \vec{v}_j\vec{v}_j^T$. This modification to the gradient computation is the only change needed by our algorithm (in Section \ref{sec:newton}) to handle this hybrid setting.
}

\subsection{The Optimization Algorithm}\label{sec:newton}
We use Newton's Method with Conjugate Gradient approximation to the Hessian to solve Problem \ref{problem:app} (Algorithm \ref{alg:newton}).
For a given value of $t_1$ and $t_2$, we use an iterative algorithm to solve the optimization problem in Equation \ref{eq:probapprox} to convergence to get an intermediate value $\mat{Z}_{t_1, t_2}$. Then, starting from $\mat{Z}_{t_1,t_2}$, we again iteratively solve Equation \ref{eq:probapprox} but with a larger value of $t_1$ and $t_2$. We repeat this process until convergence. 
Each sub-problem uses Conjugate Gradient \cite{NoceWrig06} to find an update direction for $\mat{\Sigma}$ without materializing the large Hessian matrix. We then use backtracking line search to find a step size for this direction that ensures that the updated $\mat{\Sigma}$ improves over the previous iteration and is still positive definite.


\begin{algorithm}
    \KwIn{Basis matrix $\mat{B}$, reconstruction matrix $\mat{L}$, accuracy constraints $\vec{c}$;}
    \KwOut{Solution $\mat{\Sigma}$;}
    Initialize $\mat{\Sigma} = \mat{\Sigma}_0$, $t_1=1$, $t_2=1$\;
    \For{$iter=1$ \KwTo MAXITER}{
     $\vec{s}$ = ConjugateGradient($\mat{\Sigma}, \mat{B}, \mat{L}, \vec{c}$)\;\label{line:direction}
    $\delta=\langle \vec{s},\nabla F(\mat{\Sigma}) \rangle$\; \label{line:decrease}
    \tcp{Stopping Criteria}
    \If{$|\delta| < $ NTTOL }{
    $gap = (d+m)/t_1$\;
    \If{$gap < $ TOL}
    { break\;
    }
    $t_1 = MU*t_1, \quad t_2 = MU*t_2$\; \label{line:next}
    }
    $\alpha$ = LineSearch($\mat{\Sigma}, \vec{v}$)\; \label{line:search}
    $\mat{\Sigma} = \mat{\Sigma} + \alpha * \vec{s}$\;\label{line:update}
    }
    \textbf{Return }$\mat{\Sigma}$
    \caption{Optimize($\mat{B}, \mat{L}, \vec{c}$)}
    \label{alg:newton}
\end{algorithm}

In Algorithm \ref{alg:newton}, MAXITER is the maximum number of iterations to use, while NTTOL, TOL are tolerance parameters. Larger tolerance values make the program stop faster at a slightly less accurate solution. $MU$ is the factor we use to rescale $t_1$ and $t_2$.  Typical values we used are $MU \in \{2, 5, 10 \}$. In Line \ref{line:direction}, we get an approximate Newton direction and then we use Line \ref{line:decrease} to see if this direction can provide sufficient decrease (here $\nabla F$ is the gradient of the objective function). If not, the sub-problem is over and we update $t_1$ and $t_2$ (Line \ref{line:next}) to continue with the next sub-problem. Otherwise, we find a good step size for the search direction (Line \ref{line:search}) and then update $\mat{\Sigma}$ (Line \ref{line:update}).

\subsubsection{Gradient and Hessian Computation}
The gradient and Hessian of the objective function $F$ can be derived in closed form.
Let $g_i(\mat{\Sigma}) = \vec{b}_i^T \mat{\Sigma}^{-1} \vec{b}_i$, $G_i(t_1, \mat{\Sigma}) = \exp \left( t_1 *\vec{b}_i^T \mat{\Sigma}^{-1} \vec{b}_i \right)$, $h_j(\mat{\Sigma}) = \frac{\vec{v}_j \mat{\Sigma} \vec{v}_j^T} { \vec{c}_j}$, $H_j(t_2, \mat{\Sigma}) = \exp \left(t_2*\frac{\vec{v}_j \mat{\Sigma} \vec{v}_j^T} { \vec{c}_j} \right)$. Noting that the matrix $\mat{\Sigma}$ is symmetric, we can calculate the gradient and Hessian of the functions $f_i$ and $g_i$ as follows:
\begin{align}
    \nabla g_i (\mat{\Sigma}) &= -\mat{\Sigma}^{-1} \vec{b}_i \vec{b}_i^T \mat{\Sigma}^{-1} \\
    \nabla^2 g_i (\mat{\Sigma}) &= -\nabla g_i (\mat{\Sigma}) \otimes \mat{\Sigma}^{-1} - \mat{\Sigma}^{-1} \otimes \nabla g_i(\mat{\Sigma}) \\
    \nabla h_j (\mat{\Sigma}) &= \frac{\vec{v}_j^T \vec{v}_j}{c_j} \\
    \nabla^2 h_j (\mat{\Sigma}) &= \mat{0}
\end{align}
Then the gradient and Hessian of the objective function $F(\mat{\Sigma})$ can be calculated as follows.
\begin{align}
    \nabla F(\mat{\Sigma}) &= \frac{\sum_i \left( G_i (t, \mat{\Sigma}) \nabla g_i (\mat{\Sigma}) \right)}{\sum_i G_i (t, \mat{\Sigma})} + \frac{\sum_j \left( H_j(t_2, \mat{\Sigma}) \nabla h_j(\mat{\Sigma}) \right)}{\sum_j H_j (t_2, \mat{\Sigma})}
\end{align}

\begin{align}
    \nonumber 
    \nabla^2 F(\mat{\Sigma}) &= \frac{t_1 \sum_i \left( G_i (t_1, \mat{\Sigma}) \left( \nabla^2 g_i (\mat{\Sigma}) + \nabla g_i(\mat{\Sigma}) \otimes \nabla g_i(\mat{\Sigma}) \right)\right)}{\sum_i G_i (t_1, \mat{\Sigma})} \\
    \nonumber
    &- \frac{t_1 \sum_i \left( G_i (t_1, \mat{\Sigma}) \nabla g_i (\mat{\Sigma}) \right) \otimes \sum_i \left( G_i (t_1, \mat{\Sigma}) \nabla g_i (\mat{\Sigma}) \right)}{\left( \sum_i G_i (t_1, \mat{\Sigma}) \right)^2} \\
    \nonumber
    &+ \frac{t_2 \sum_j \left( H_j(t_2, \mat{\Sigma})\left( \nabla h_j(\mat{\Sigma}) \otimes \nabla h_j(\mat{\Sigma}) \right)\right)}{\sum_j H_j (t_2, \mat{\Sigma})} \\
    &- \frac{t_2 \sum_j \left( H_j(t_2, \mat{\Sigma}) \nabla h_j(\mat{\Sigma}) \right) \otimes \sum_j \left( H_j(t_2, \mat{\Sigma}) \nabla h_j(\mat{\Sigma}) \right)}{\left( \sum_j H_j (t_2, \mat{\Sigma}) \right)^2 }\label{eq:thehessian}
\end{align}
Here $\otimes$ is the Kronecker product. Because of the Kronecker product, multiplication of the Hessian by a search direction can be done without materializing the Hessian itself. Specifically, we use the well-known property $\left(\mat{A} \otimes \mat{B}\right) vec\left( \mat{C} \right) = vec \left( \mat{B} \mat{C} \mat{A}^T\right)$ that is frequently exploited in the matrix mechanism literature \cite{yuan2015optimizing,mckenna2018optimizing}. We let HessTimesVec denote the function that exploits this trick to efficiently compute the multiplication $\mat{H}\vec{p}$ of the Hessian (of objective function $F$) by a search direction without explicitly computing $\mat{H}$.

\begin{algorithm}
    \KwIn{Variable $\mat{\Sigma}$, basis matrix $\mat{B}$, index matrix $\mat{L}$, accuracy constraints $\vec{c}$;}
    \KwOut{Search direction $\vec{s}$;}
    Initialize $\vec{s}=0$, $\vec{r}=- \nabla F(\mat{\Sigma})$, $\vec{p}=\vec{r}$, $rsold=\langle \vec{r}, \vec{r}\rangle$ \;
    \For{$i=1$ \KwTo MAXCG}{\label{line:maxcg}
    $\vec{Hp}$ = HessTimesVec($\vec{p}$) \;
    $a=\frac{rsold}{\vec{p}^T*\vec{Hp}}$, $\vec{s}=\vec{s}+a*\vec{p}$,  $\vec{r}=\vec{r}-a*\vec{Hp}$, $rsnew = \langle \vec{r}, \vec{r} \rangle$ \;
    \If{$rsnew \leq $TOL2}{\label{line:terminatecg}
    break\;
    }
    $b=\frac{rsnew}{rsold}$, $\vec{p} = \vec{r}+ b*\vec{p}$, $rsold=rsnew$\;
    }
    \textbf{Return} $\mat{s}$
    \caption{ConjugateGradient($\mat{\Sigma}, \mat{B}, \mat{L}, \vec{c}$)}
    \label{alg:direction}
\end{algorithm}

\subsubsection{Conjugate Gradient}
Algorithm \ref{alg:newton} finds a search direction using the Conjugate Gradient algorithm, which is commonly used for large scale optimization \cite{NoceWrig06}. The main idea is that Newton's method uses second-order Taylor expansion to approximate $F(\mat{\Sigma})$ and would like to compute the search direction  by solving
\begin{align}
    \vec{s} = \arg \min_{\vec{s}} \frac{1}{2} \vec{s}^T \nabla^2 F(\mat{\Sigma}) \vec{s} + \vec{s}^T \nabla F(\mat{\Sigma}) 
    \label{problem:taylor}
\end{align}
whose solution is $-\mat{H}^T\nabla F$, where $H=\nabla^2 F$ is the Hessian.
However, the size of Hessian matrix is $d^2\times d^2$ (where $d$ is the size of the domain of possible tuples in our case). This is intractably large, but fortunately, only approximate solutions to Equation \ref{problem:taylor} are necessary for optimization, and this is what conjugate gradient does \cite{NoceWrig06}. Algorithm \ref{alg:direction} provides the pseudocode for our application (recall that HessTimesVec is the function that efficiently computes the product of the Hessian times a vector by taking advantage of the Kronecker products in Equation \ref{eq:thehessian}).
We note that Yuan et al. \cite{yuan2015optimizing} used 5 conjugate gradient iterations in their matrix mechanism application. Similarly we use 5 iterations (MAXCG = 5) in Line \ref{line:maxcg}. We also terminate the loop early if there is very little change, checked in Line \ref{line:terminatecg} (we use  TOL2 $ = 10^{-10}$).

\subsubsection{Step Size}
Once the conjugate graident algorithm returns a search direction, we need to find a step size $\alpha$ to use to update $\mat{\Sigma}$ in Algorithm \ref{alg:newton}. For this we use the standard backtracking line search \cite{NoceWrig06}, whose application to our problem is shown in Algorithm \ref{alg:step}. It makes sure that the step size is small enough (but not too small) so that it will (1) result in a positive definite matrix and (2) the objective function will decrease sufficiently. We use Cholesky decomposition to check for positive definiteness.
%
The parameter $\sigma$ determines how much the objective function need to decrease before breaking the iteration, a typical setting is $\sigma = 0.01$. 
\begin{algorithm}
    \KwIn{Variable $\mat{\Sigma}$, search direction $\vec{s}$;}
    \KwOut{Step size $\alpha$;}
    $fcurr=F(\mat{\Sigma})$, $flast=fcurr$, $\mat{\Sigma}_{old} = \mat{\Sigma}$, $j=1$, $\beta=0.5$\;
    \While{true}{
    $\alpha=\beta^{j-1}$, $j=j+1$, $\mat{\Sigma}_{new} = \mat{\Sigma}_{old}+\alpha*\vec{s}$, $fcurr=F(\mat{\Sigma}_{new})$ \;
    \If{ $ \mat{\Sigma}_{new} \preceq 0$ }{
    continue\;
    }
    \If{$fcurr\leq flast+\alpha *\sigma *\langle \vec{s}, \vec{\nabla F(\mat{\Sigma})} \rangle$}{
    break\;
    }
    }
    \textbf{Return} $\alpha$
    \caption{LineSearch($\mat{\Sigma}, \vec{s}$)}
    \label{alg:step}
\end{algorithm}

\subsection{Initialization}\label{sec:init}
Optimization algorithms need a good initialization in order to converge reasonably well. However, specifying a correlation structure for Gaussian noise is typically not an intuitive approach for data publishers. Instead, data publishers may feel more comfortable specifying a query matrix $\mat{Q}$ (with linearly independent rows) to which it may be reasonable (as a first approximation) to add independent noise. That would result in an initial suboptimal mechanism $M_0(\vec{x}) = \mat{W}\mat{Q}^+(\mat{Q}\vec{x} + N(0, \sigma^2\mat{I}))$ \cite{LMHMR15}, where $\mat{Q}^+$ is the Moore-Penrose pseudo-inverse of $\mat{Q}$ \cite{GoluVanl96} -- that is, this suggested mechanism would add independent noise to $\mat{Q}\vec{x}$ and then recover workload query answers by multiplying the result by $\mat{W}\mat{Q}^+$.

We do not run this mechanism, instead we derive a Gaussian correlation $\mat{\Sigma}$ from it. The covariance matrix for that mechanism would be 
$\sigma^2\mat{W}\mat{Q}^+(\mat{W}\mat{Q}^+)^T$. Noting that $\mat{W}=\mat{L}\mat{B}$, this can be written as $\sigma^2\mat{L}\mat{B}\mat{Q}^+(\mat{L}\mat{B}\mat{Q}^+)^T$. This is equivalent to adding $N(\vec{0}, \sigma^2\mat{Q}^+(\mat{Q}^+)^T)$ noise to $\mat{B}\vec{x}$ and hence we can set the initial $\mat{\Sigma}$ to be $\sigma^2\mat{Q}^+(\mat{Q}^+)^T$. 

Next we need to choose a small enough value for $\sigma^2$ so that the initial covariance matrix $\mat{\Sigma}=\sigma^2\mat{Q}^+(\mat{Q}^+)^T$ would satisfy the fitness-for-use constraints. We do this by choosing

%
%
%
\begin{align*}
\sigma^2 = \min_{j=1\cdots d} \frac{c_j}{j^\text{th}\text{ element of }diag\left(  \mat{L}\mat{Q}^+ \left(\mat{Q}^+ \right)^T\mat{L}^T \right) }*0.99
\end{align*}
Our optimizer (Algorithm \ref{alg:newton}) starts with this $\mat{\Sigma}$ and iteratively improves it. 

We can also offer guidance about the setting for $\mat{Q}$ in cases where the data publisher does not have a good guess. A simple strategy is to set $\mat{Q}=\mat{I}$; in fact, this is the setting we use for our experiments (to show that our algorithms can succeed without specialized knowledge about the problem). It is also possible to set $\mat{Q}$ to be the result of Matrix Mechanism algorithms such as \cite{yuan2015optimizing}. 


\section{Experiments}\label{sec:experiments}
In these experiments, our proposed algorithm is referred to as \algoname~ (for the two applications of soft max). \addcolor{We compare against the following algorithms.} \addcolor{Input Perturbation (\gmname) adds independent Gaussian noise directly to the data vector}. The optimal \addcolor{Matrix Mechanism Convex Optimization Algorithm (CA) \cite{YYZH16} minimizes sum-squared error under  $(\epsilon,\delta)$-differential privacy}.
\addcolor{Since queries can have different relative importance, we also consider two variations of CA. \wcaname weights each query by the inverse of its target variance (so queries that need low variance have higher weight in the objective function) and \wcainame weights each query by the inverse of the square root of the variance.}
We also consider the Hierarchical Mechanism (HM) \cite{hay2009boosting,hbtree} as it is good for some workloads. Specifically, HM uses a strategy matrix $\mat{H}$ which represents a tree structure with \addcolor{optimal} branching factor \cite{hbtree}. 
\addcolor{We also considered the Gaussian Mechanism (GM), which adds independent Gaussian noise to the workload query answers. However, it generally performed worse than IP and always significantly worse than CA, so we dropped it from the tables to save space.}
We do not  compare with alternative Matrix Mechanism algorithms like Low-Rank Mechanism \cite{YZWXYH12}, Adaptive Mechanism \cite{li2012adaptive} and Exponential Smoothing Mechanism \cite{YZWXYH12} because the CA baseline is optimal for Matrix Mechanism \cite{YYZH16} under $(\epsilon,\delta)$-differential privacy. 

We compare these algorithms on a variety of workloads \addcolor{based on their performance at the same privacy cost}. Our algorithm SM-II finds the minimal privacy cost needed to satisfy the accuracy constraints. We then set the other algorithms to use this privacy cost and we check by how much they exceed the pre-specified variance requirements (i.e., to what degree are they sub-optimal for the fitness-for-use problem). Note that these algorithms are all data-independent, so the variance can be computed in closed form and will be the same for any pair of datasets that have the same \addcolor{dimensionality and domain size}. \addcolor{One can also compare privacy cost needed to reach the target variance as follows. If, given the same privacy cost, Method 1 exceeds its target variance by a factor of $x_1$ and Method 2 exceeds it by $x_2$ (which is what our experiments show), then if they were both given enough privacy cost to meet their target variance goals, the ratio of privacy cost for Method 1 to that of Method 2 is $\sqrt{x_1/x_2}$ (this follows from Theorem \ref{thm:equiv}).}

We follow Section \ref{sec:init} to initialize our algorithm by setting the initialization parameter $\mat{Q}=\mat{I}$ (the $d\times d$ identity matrix). For basis matrices $\mat{B}$, we use two choices: $\mat{B}=\mat{I}$ or $\mat{B}=\mat{U}$, where $\mat{U}$ is an upper triangle matrix ($\mat{U}[i, j] = 1$ if $i\leq j$ and $\mat{U}[i, j] = 0$ otherwise).

%
All experiments are performed on a machine with an Intel i7-9750H 2.60GHz CPU and 16GBytes RAM.

\subsection{Evaluating Design Choices}
We made two key design choices for our algorithm. The first was to approximate the constrained optimization in Problem \ref{problem1} with the soft max optimization in Problem \ref{problem:app}. Then we used a Newton-style method to solve it instead of something simpler like gradient descent. We evaluate the effect of these choices here.

\subsubsection{The Soft Max Approximation}\label{sec:exp:approx}
For small problems, we can directly solve Problem \ref{problem1} by implementing an interior point method \cite{boyd2004convex} and so we can compare the resulting privacy cost with the one returned by our soft max approximation \algoname. For this experiment, we considered prefix-range queries over an ordered domain of size $d$. These are one-sided range queries (i.e., the sum of the first 2 records, the sum of the first 3 records) from which all one-dimensional range query answers can be computed by subtraction of two one-sided range queries. Thus the workload $\mat{W}$ is a lower triangular matrix where the lower triangle consists of ones. The result is shown in Table \ref{table:pcost_pr} and shows excellent agreement. Note that $d=64$ is the limit for our interior point solver, while with \algoname we can easily scale to $d=1024$ on a relatively weak machine (using a server-grade machine or a GPU can improve scalability even more). We note that  extreme scalability, like in the high-dimensional matrix mechanism \cite{mckenna2018optimizing}, is an area of future work and we believe that a combination of the ideas from  \cite{mckenna2018optimizing} and our work can make this possible, by breaking a large optimizaton problem into smaller pieces and using our \algoname method to optimize each smaller piece.

\begin{table}
\caption{Privacy cost comparison between \algoname and interior point method. Note that IP did not scale well, thus limiting the size of $d$.}
\centering
 \begin{tabular}{| c | c | c | c | c | c |  } 
 \hline
 d & 2 & 4 & 8 & 16 & 64  \\ [0.5ex] 
  \hline
 Interior Point & 1.33 &  1.76 & 2.28 & 2.91  &4.46 \\
  \hline
 \algoname & 1.33 &  1.76 & 2.28 & 2.91 &4.46 \\
 \hline
\end{tabular}

\label{table:pcost_pr}
\end{table}

\subsubsection{\algoname vs. Gradient Descent}
As far as optimizers go, \algoname is relatively simple, but not as simple as the very popular gradient descent (GD). To justify the added complexity, we compared these two optimizers for solving Problem \ref{problem:app}. We used the same setup as in Section \ref{sec:exp:approx} and the results are shown in Table \ref{table:time_gd}. Clearly the use of \algoname is justified as gradient descent does not converge fast enough even at small problem sizes.
\begin{table}
\caption{Convergence time for \algoname vs. Gradient Descent (GD) in seconds. Even at small scales, gradient descent is vastly inferior.}
\centering
 \begin{tabular}{| c | c | c | c | c |  } 
 \hline
 d & 2 & 4 & 8 & 16  \\ [0.5ex] 
  \hline
 GD & 7.02 & 44.90 & 198.52 & 264.36  \\
  \hline
 \algoname & 0.027 & 0.055 & 0.053 & 0.058 \\
 \hline
\end{tabular}

\label{table:time_gd}
\end{table}

\subsection{The Identity-Sum Workload}
\addcolor{
In Section \ref{sec:toy}, we considered the workload consisting of the identity query and the sum query, with equal target variance. We further explore this example to illustrate further surprising behavior of algorithms that optimize for sum-squared-error (so for just this experiment we focus on CA, \wcaname, \wcainame). Specifically, it is believed that in sum-squared error optimization, weighting queries by the inverse of their desired variance helps improve their individual accuracy. We show that this is not always the case.}

\addcolor{In these experiments, the identity query has target variance 1, but we will vary the sum query's target variance, which we denote by the variable $k$.}

\begin{figure}
\centering
\includegraphics[clip, trim=2cm 9cm 2cm 9cm, width=0.9\columnwidth]{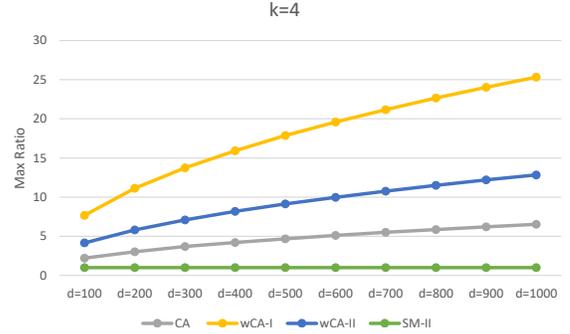}
\caption{Comparison for different $d$ when  $k=4$.}\label{fig:k=4}
\end{figure}

%
%
\addcolor{In Figure \ref{fig:k=4}, we set $k=4$ and vary the domain size $d$. As expected, \algoname outperforms
the others. What is surprising is that \wcaname and \wcainame perform worse than CA because this means that the typical recommended weighting strategy actually makes things worse (the sum query exceeds its target variance by a larger ratio). For reference, we derive the optimal analytical solutions for these methods on this workload in the appendix\ifbool{ARXIV}{.}{ of the full version of our paper \cite{fullfitness}.}}


\begin{figure}
    \centering
    \includegraphics[clip, trim=2cm 9.2cm 2cm 9.2cm, width=0.9\columnwidth]{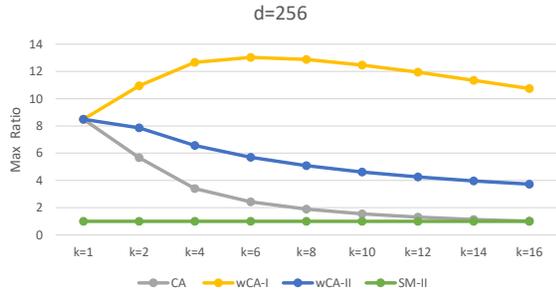}
    \caption{Comparison for different $k$ when $d=256$.}
    \label{fig:n=256}
\end{figure}

\addcolor{Figure \ref{fig:n=256} fixes $d$ at 256 and varies $k$ (starting from $k=1$). Again we see the same qualitative effects as in Figure \ref{fig:k=4}.}

\subsection{PL-94}
PL94-171 \cite{pl94} is a Census dataset used for redistricting. In 2010, it contained the following variables: 
\begin{itemize}[leftmargin=*]
    \item \textbf{voting-age}: a binary variable where $0$ indicates someone is 17 years old and under, while $1$ indicates 18 or over.
    \item \textbf{ethnicity}: a binary variable used to indicate if someone is Hispanic (value $1$) or not (value $0$).
    \item \textbf{race}: this is a variable with 63 possible values. The OMB race categories in this file in 2010 were \emph{White}, \emph{Black or African American}, \emph{American Indian and Alaska Native}, \emph{Asian}, \emph{Native Hawaiian and Other Pacific Islander}, and \emph{Some Other Race}. An individual is able to select any non-empty subset of races, giving a total of 63 possible values.
\end{itemize}
These variables create a (2, 2, 63) histogram which can be flattened into a 252-dimension vector.

For this experiment, we created a workload consisting of multiple marginals. These were a) voting-age marginal (i.e., number of voting-age people and number of non-voting-age people); b) ethnicity marginal; c) the number of people in each OMB category who selected only one race; d) for each combination of 2 or more races, the number of people who selected that combination; e) the identity query (i.e., for each demographic combination, how many people fit into it). 
\addcolor{As we are not aware of any public error targets, we will assume all queries are equally important and set all variance targets to 1.} 
The basis matrix is $\mat{B} = \mat{U}$ (upper triangular matrix).


We found the minimal privacy cost needed to match these variance bounds using \algoname and set that as the privacy cost for all algorithms. Then we measured the maximum ratio of query variance to target variance for each method. The results are shown in
Table \ref{table:pl_compare}. We see that  \addcolor{input perturbation performs the worst}. The matrix mechanism optimized for total error (CA) is much better, but it still has a maximum variance ratio of  \addcolor{3.99} times larger than optimal. This means that the matrix mechanism is more accurate than necessary for some queries at the expense of missing the target bound for other queries. Meanwhile, \algoname can match the desired bounds for each query, which is what it is designed to do. \addcolor{In terms of \emph{sum squared error}, which is what CA is optimized for, the error for SM-II is 2.07 times that of CA.}



\begin{table}
\caption{Max variance ratio on PL94 with uniform targets.}
\centering
 \begin{tabular}{| c | c | c | c | c | c| c|}
 \hline
 Mechanism & \addcolor{\gmname} & HM & CA & \addcolor{\wcaname}  &\addcolor{\wcainame} &\algoname \\ [0.5ex] 
  \hline
 PL94 &36.56	&13.93	&3.99 & 3.99 &3.99	&1.00
 \\
  \hline
\end{tabular}
\label{table:pl_compare}
\end{table}

\subsection{Range Queries}
We next consider one-dimensional range queries. Here we follow a similar setting to \cite{YZWXYH12}. We vary $d$, the number of items in an ordered domain. The workload is composed of $2*d$ random range queries, where the end-points of each query are sampled uniformly at random. We compare two settings, (1) the uniform cases where the target variance bound is equal to $1$ for each query, and (2) the random case where the target variance bound for each query is randomly sampled from a uniform$(1,10)$ distribution. In the experiments we vary the domain size $d \in \{ 64,128, 256,512, 1024\}$, and use the basis matrix $\mat{B} = \mat{U}$ (upper triangular matrix).
 
Again, we find the minimum privacy cost using \algoname, set each algorithm to use that privacy cost and then compute the maximum variance to target variance ratio for each method. The results for the uniform targets are shown in Table \ref{table:compare}\addcolor{, with \algoname achieving $\approx 25\%$ improvement over the best competitor.} The results for the random targets are shown in Table \ref{table:compare_random}\addcolor{, with \algoname improving by at least 40\% over competitors.} \addcolor{In the uniform case, the \emph{sum squared error} of SM-II ranges from 1.04 to 1.09 times that of  CA, depending on the value of $d$.}
 
%


\begin{table}
\caption{Maximum ratio of achieved variance to target variance on Range Queries with uniform accuracy targets.}
\centering
 \begin{tabular}{| c | c | c | c | c | c| c|}
 \hline
 Mechanism &  \addcolor{\gmname} & HM & CA & \addcolor{\wcaname} & \addcolor{\wcainame}& \algoname \\ [0.5ex] 
  \hline
 d = 64 &10.63	&2.84	&1.27 &1.27	&1.27	&1.00
 \\
  \hline
   d = 128 &17.45	&3.48	&1.25 	&1.25	&1.25	&1.00
 \\
  \hline
   d = 256 &28.43	&3.90	&1.29 &1.29	&1.29&1.00
 \\
  \hline
   d = 512 &47.36	&4.68	&1.23	&1.23	&1.23	&1.00
 \\
  \hline
   d = 1024 &77.26	&5.42	&1.24	&1.24 &1.24	&1.00
 \\
  \hline
\end{tabular}
\label{table:compare}
\end{table}

\begin{table}
\caption{Maximum ratio of achieved variance to target variance on Range Queries with random accuracy targets.}
\centering
 \begin{tabular}{| c | c | c | c | c | c| c|}
 \hline
 Mechanism &\addcolor{\gmname} & HM & CA & \addcolor{\wcaname} & \addcolor{\wcainame}& \algoname \\ [0.5ex] 
  \hline
 d = 64 & 16.04	& 4.63	& 2.53 & 1.62 &1.98  &1.00
 \\
  \hline
   d = 128 &21.28	& 5.87	&2.14 & 1.49&1.75	&1.00
 \\
  \hline
   d = 256 &41.76	&6.05	&1.91 & 1.45&1.69	&1.00
 \\
  \hline
    d = 512 & 71.51 	& 7.35  & 1.80 &  1.46&1.52	&1.00
 \\
  \hline
   d = 1024 &109.34 	& 7.68	& 1.69 & 1.43&1.49	&1.00
 \\
  \hline
\end{tabular}
\label{table:compare_random}
\end{table}




\begin{table}
\caption{Maximum achieved variance to target variance ratio on Random Queries with a Uniform target variance.}
\centering
 \begin{tabular}{| c | c | c | c | c | c| c|}
 \hline
 Mechanism & \addcolor{\gmname} & HM & CA & \addcolor{\wcaname} & \addcolor{\wcainame}& \algoname\\ [0.5ex] 
  \hline
 d = 64 &1.14	&2.29	&1.14 &1.14 &1.14	&1.00
 \\
  \hline
   d = 128 &1.14	&2.29	&1.07 &1.07	 &1.07&1.00
 \\
  \hline
   d = 256 &1.14	&3.39	&1.06 &1.06	 &1.06&1.00
 \\
  \hline
   d = 512 &1.15	&3.40	&1.05 &1.05	 &1.05&1.00
 \\
  \hline
   d = 1024 &1.15	&3.41	&1.04 &1.04	  &1.04&1.00
 \\
  \hline
\end{tabular}
\label{table:compare-dis}
\end{table}

%


\begin{table}
\caption{Maximum achieved variance to target variance ratio on Random Queries with a Random target variance.}
\centering
 \begin{tabular}{| c | c | c | c | c | c| c| }
 \hline
 Mechanism &\addcolor{\gmname} & HM & CA & \addcolor{\wcaname} & \addcolor{\wcainame}& \algoname \\ [0.5ex] 
  \hline
 d = 64 &2.58 &7.44 &2.43 &1.15 &1.61 &1.00
 \\
  \hline
   d = 128  &2.82 &8.28 &2.58 &1.18 &1.72 &1.00
 \\
  \hline
   d = 256  &2.82 &8.36 &2.58 &1.17 &1.70 &1.00
 \\
  \hline
   d = 512  &2.79 &8.27 &2.51 &1.16 &1.68 &1.00
 \\
  \hline
   d = 1024  &2.76 &8.17 &2.49 &1.16 &1.67 &1.00
 \\
  \hline
\end{tabular}
\label{table:compare-dis-random-2}
\end{table}

\subsection{Random Queries}
We next consider a random query workload similar to the setting in \cite{YZWXYH12}. Here for each entry in each $d$-dimensional query vector, we flip a coin with $P(\text{heads})=0.2$. If it lands heads, the entry is set to 1, otherwise it is set to $-1$. Again we consider two scenarios. \addcolor{(1) the uniform cases where the target variance bound is equal to $1$ for each query, and (2) the random case where the target variance bound for each query is randomly sampled from a uniform$(1,10)$ distribution.} We vary $d \in \{ 64,128, 256,512, 1024\}$, and the number of queries is $m=2*d$. The basis matrix is $\mat{B} = \mat{I}$.


Table \ref{table:compare-dis} lists the maximum variance ratio among all queries under the same privacy cost for different methods for the uniform scenario. \addcolor{Table \ref{table:compare-dis-random-2} shows the corresponding results for random targets. In both cases, the best competitor}
 is not much worse than \algoname. We speculate that this is because the queries do not have much structure that can be exploited to reduce privacy cost. \addcolor{In terms of \emph{sum squared error}, for the case of uniform target variances, the total error of SM-II was at most $1.005$ times that of CA.}

\begin{table}
\caption{Maximum achieved variance to target variance ratio for Age pyramids with uniform accuracy targets.}
\centering
 \begin{tabular}{| c | c | c | c | c | c| c|  }
 \hline
 Mechanism & \addcolor{\gmname} & HM & CA & \addcolor{\wcaname}  &\addcolor{\wcainame} &\algoname \\ [0.5ex] 
  \hline
 AGE &32.49	&6.2	&1.58	&1.58 &1.58  &1.00
 \\
  \hline
\end{tabular}
\label{table:age_compare_2}
\end{table}

\subsection{Age Pyramids}
\addcolor{Demographers often} study the distribution of ages in a population by gender. We consider a dataset schema similar to the Census, where gender is a binary attribute and there are 116 ages (0-115). 
The range queries we consider are prefix queries (i.e., age $\in[0, x]$ for all $x$) and age $\in[18,115]$ \addcolor{(the voting age population)}. The workload consists of (1) range queries for males, (2) range queries  for females, (3) range queries \addcolor{ for all. We set uniform target variance bounds of 1.} We use the basis matrix $\mat{B} = \mat{U}$ (upper triangular).
%
%
Table \ref{table:age_compare_2} shows the maximum  ratio of achieved variance to target variance for each algorithm, with \algoname clearly outperforming the competitors \addcolor{by at least 58\%}. \addcolor{Meanwhile the sum squared error of SM-II is 1.07 times that of CA.}




\subsection{Marginals}
\addcolor{We next consider histograms $H$ on \addcolor{$r$} variables, where each variable can take on one of \addcolor{$t$} values, resulting in a domain size of \addcolor{$d=t^r$}. 
We consider the workload consisting of all one-way and two-way marginals with uniform target variance of 1.}
%
\addcolor{For this set of experiments, we set $r=3$ and varied the  values of $t$.
We used the basis matrix is $\mat{B} = \mat{I}$ for \algoname. 
\addcolor{Table \ref{tab:marginal} shows the maximum target variance ratio of different algorithms.}  \algoname outperforms the competitors, especially as $t$ increases.}
\begin{table}
\caption{Maximum achieved variance to target variance ratio on Marginal Queries when $r=3$}
\centering
 \begin{tabular}{| c | c | c | c | c | c| c|}
 \hline
 Mechanism & \addcolor{\gmname} & HM & CA & \addcolor{\wcaname}  &\addcolor{\wcainame} &\algoname \\ [0.5ex] 
  \hline
 t = 2      &1.82	&3.02	&1.14	&1.14	&1.14	&1.00\\
  \hline
   t = 4    &4.55	&10.28	&1.42	&1.42	&1.42	&1.00\\
  \hline
   t = 6    &8.6	&21.38	&1.66	&1.66	&1.66	&1.00   \\
  \hline
    t = 8    &14.03	&36.72	&1.88	&1.88	&1.88	&1.00\\
  \hline
    t = 10   &20.84	&56.23	&2.08	&2.08	&2.08	&1.00 \\
  \hline
  t = 12   &28.76	&78.96	&2.25	&2.25	&2.25	&1.00 \\
  \hline
   t = 14  &38.17  &106.22  &2.41 &2.41 &2.41 &1.00\\
  \hline
     t = 16  &48.85  &137.41  &2.57 &2.57 &2.57 &1.00\\
  \hline
\end{tabular}
\label{tab:marginal}
\end{table}
\addcolor{The ratio of sum squared error of SM-II to CA ranged from 1.1 ($t=2$) to 1.34 ($t=16$).}

\subsection{WRelated Queries}
\addcolor{We next consider the one-dimensional WRelated workload of \cite{YZWXYH12}. The workload matrix is $W = C A$. Here matrix $C$ has size $m \times s$ and matrix $A$ has size $s \times d$, each follows the Gaussian(0, 1) distribution. In the experiment we set $m = d/2, s = d/2$. Due to space constraints, we only show results for the uniform target variance bound of 1.
We vary the domain size $d \in \{ 64,128, 256,512, 1024\}$, and use the basis matrix $\mat{B} = \mat{I}$.
%
The results in Table \ref{table:related_uniform} shows \algoname  outperforming competitors by around 31\% on the smaller domain sizes. The ratio of total  squared error for SM-II to that of CA was at most 1.08 }

\begin{table}
\caption{Maximum achieved variance to target variance ratio on WRelated Queries with a Uniform target variance.}
\centering
 \begin{tabular}{| c | c | c | c | c | c| c|}
 \hline
 Mechanism & \addcolor{\gmname} & HM & CA & \addcolor{\wcaname} & \addcolor{\wcainame}& \algoname\\ [0.5ex] 
  \hline
 d = 64 &3.51	&9.59	&1.31 &1.31 &1.31	&1.00
 \\
  \hline
  d = 128 &4.16	&11.16	&1.31 &1.31 &1.31	&1.00
 \\
  \hline
  d = 256 &4.28	&12.60	&1.19 &1.19 &1.19 &1.00
 \\
  \hline
  d = 512 &3.62	&10.46	&1.12 &1.12 &1.12	&1.00
 \\
  \hline
  d = 1024 &3.50	&10.05	&1.07 &1.07 &1.07	&1.00
 \\
  \hline
\end{tabular}
\label{table:related_uniform}
\end{table}


\section{Conclusions and Future Work}\label{sec:conc}
In this work we introduce the fitness-for-use problem, where the goal is to calculate minimal privacy cost under accuracy constraints for each query. \addcolor{After theoretical analysis of the problem, we proposed an algorithm named \algoname to solve it}. While our algorithm used variance as the accuracy measure for each query, other applications may require their own specific accuracy measures. This consideration leads to two important directions for future work.
The first is to create optimized differentially private algorithms that meet fitness-for-use goals under measures other than squared error. The second direction is to achieve the same kind of extreme scalability as the high-dimensional matrix mechanism \cite{mckenna2018optimizing} that provided a tradeoff between optimality and scalability.

\section*{Acknowledgments}\label{sec:ack}
This work was supported by NSF Awards CNS-1702760 and CNS-1931686. We are grateful to Aleksandar Nikolov for discussions about factorization mechanisms.

\bibliographystyle{ACM-Reference-Format}
\bibliography{ref2.bib}

\ifbool{ARXIV}{
\clearpage
\appendix
\section{Proofs from Section \ref{sec:prob}}\label{app:problem}
\printProofs[section3]

\section{Proofs from Section \ref{sec:theo}}\label{app:theory}
\printProofs[section5]

\subsection{Proofs of Uniqueness for Solution Results}\label{app:unique}
\printProofs[uniqueness]

\subsection{Proofs from Section \ref{sec:optimize}}\label{app:optimize}
\printProofs[approx]

}{}
\end{document}